# Frustration in system with intercalated and bilayer smectic structure

Calum J. Gibb,[a†*] Claudio Flores Fernandez,[a] Magdalena M. Majewska,[b] John M. D. Storey,[a] Corrie T. Imrie, [a§] Ewa Gorecka,[b] Damian Pociecha,[b*]

[a] Department of Chemistry, School of Natural and Computing Sciences, University of Aberdeen, Meston Building, Aberdeen AB24 3UE, United Kingdom

[b] University of Warsaw, Faculty of Chemistry, ul. Zwirki i Wigury 101,02-089Warsaw, Poland

[†] Current Address: Department of Chemistry, University of Leeds, LS2 9JT, United Kingdom

[§] Deceased January 2025

*Authors for correspondence: e-mail c.j.gibb@leeds.ac.uk, pociu@chem.uw.edu.pl

**Abstract**

The dimeric mesogens exhibit an unusual variety of mesophases – nematic and smectic, with distinctly different packing schemes. The relation between molecular and mesophase structure was studied for homologue series of non-symmetric bent dimers CBO6.Om, obtained by serial Pd- and Pt-catalyzed cross-coupling chemical reactions. The shorter homologues of the series form smectics with an intercalated structure, and longer homologues form bilayer smectics with broken up-down symmetry of molecular orientation within the layers, for intermediate homologues frustration leads to a nematic gap. The additional structure complexity arises from the tendency of bent dimers to form twisted structures, in which the azimuthal angle of tilted molecules forms both short- and long-pitch helices. Here, we confirm, through resonant X-ray diffraction, that longer homologues form a helical four-layer structure of $SmC_{TB\text{-}SH}$ phase, along with a more complex one in which, in addition to the basic four-layer unit, an extra helix is superimposed, $SmC_{TB\text{-}DH}$ phase. This additional helix has a periodicity that is strongly temperature-dependent; upon cooling, it changes from 100 nm to 40 nm.

## Introduction

The liquid crystalline properties of symmetric dimers have been extensively studied. Their straightforward chemistry facilitates the synthesis of a wide range of materials.[1] In contrast, research on non-symmetric dimers is limited, as connecting two unlike mesogenic units typically requires more complex synthetic strategies.[2] Among the most investigated non-symmetric dimers are those with a diether specer (O*n*O), where *n* refers to the number of carbon atoms in the spacer), owing to their relatively simple organic synthesis.[3] Recently, interest in non-symmetric mesogenic dimers has shifted toward those with methylene and ether linkages (*n*O).[4] For dimers with odd number of atoms in the spacer spontaneous mirror symmetry breaking in these materials leads to formation, apart from twist bend-nematic phase [5-7] also its helical smectic analogues: $SmC_{TB}$ type phases.[8,9] However, the synthesis of such dimers poses significant challenges, as traditional chemical methods have often proven inadequate. Substantial progress has been achieved only in recent years, thanks to the development of Pd- and Pt-catalyzed cross-coupling reactions.[10-12] These methods enable the formation of carbon-carbon bonds with high chemical selectivity, allowing for the synthesis of previously inaccessible asymmetric liquid crystal dimers.

Here, we present the synthesis and characterization of the 1-(4-cyanobiphenyl-4′-yloxy)-6-(4-alkyloxyanilinebenzylidene)-4′-hexanes, which we refer to by the acronym CBO6.O*m* (**Fig. 1**), where *m* refers to the number of carbon atoms in the terminal chain. The CBO6.O*m* series are structural isomers of the previously reported CB6O.O*m* series [13,14] which exhibit extensive liquid crystal polymorphism, including the $N_{TB}$ phase and both the $SmC_{TB-SH}$ and $SmC_{TB-DH}$ phases.[9,15] The homologues of CBO6.O*m* series are the first non-symmetric dimers incorporating an odd-membered 'O*n*' spacer, synthesized through a novel pathway involving various Pd- and Pt-catalyzed cross-coupling reactions. The phase sequence for these compounds, depending on homologue number, involves intercalated or bilayer smectics, moreover spontaneously helical structure is formed in nematic but also in smectic phases.

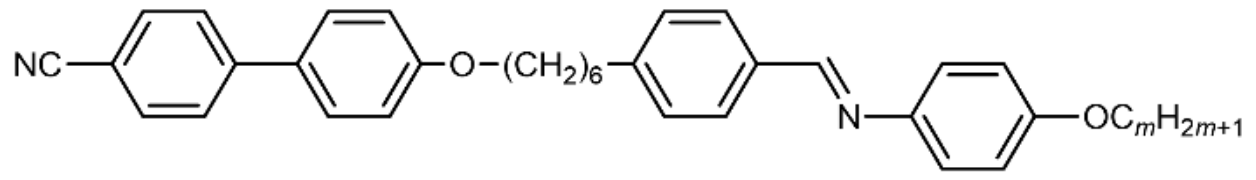

**Figure 1.** The molecular structure of the CBO6.O*m* series.

## Experimental

### *Organic Synthesis*

The synthetic route for the preparation of the CBO6.O*m* series is shown in **Scheme 1**. The route begins with a standard Williamson ether synthesis between 1-4 iodo, bromobenzene and 5-hexyne-1-ol to form 1-bromo-4-(hex-5-yn-1-yloxy)benzene (**1**). (**1**) then undergoes a Pd-catalysed Sonagasheria cross-coupling reaction with 1-4 iodobenzaldehyde using Bis(triphenylphosphine)palladium(II) dichloride as the catalyst to form 4-(6-(4'-bromophenoxy)hex-1-yn-1-yl)benzaldehyde (**2**).[11] (**2**) was then selectively hydrogenated at 4.0 bar of $H_{2(g)}$ using a PARR 5500 Series Compact Reactor hydrogenator. $Pd/TiO_2$ (0.5% *w*/*w*) was added as the catalyst to reduce the alkyne moieties yielding the 4-(6-(4'-bromophenoxy)hexyl)benzaldehyde (**3**). Over hydrogenation of (**2**) is possible using the catalyst chosen and the uptake of $H_{2(g)}$ was carefully monitored throughout the reaction. Once 2 mole equivalence of gas is consumed by the reaction, hydrogenation is stopped. (**3**) underwent a Suzuki cross-coupling reaction with 1,4 cyanophenyl boronic acid to form 4'-((6-(4-formylphenyl)hexyl)oxy)-[1,1'-biphenyl]-4-carbonitrile (**4**).[16] The final synthetic step is a standard Schiff base synthesis between the relevant 1-4 alkoxy aniline and (**4**) to form the 1-(4-cyanobiphenyl-4′-yloxy)-6-(4-alkyloxyanilinebenzylidene)-4′hexanes (CBO6.O*m*).

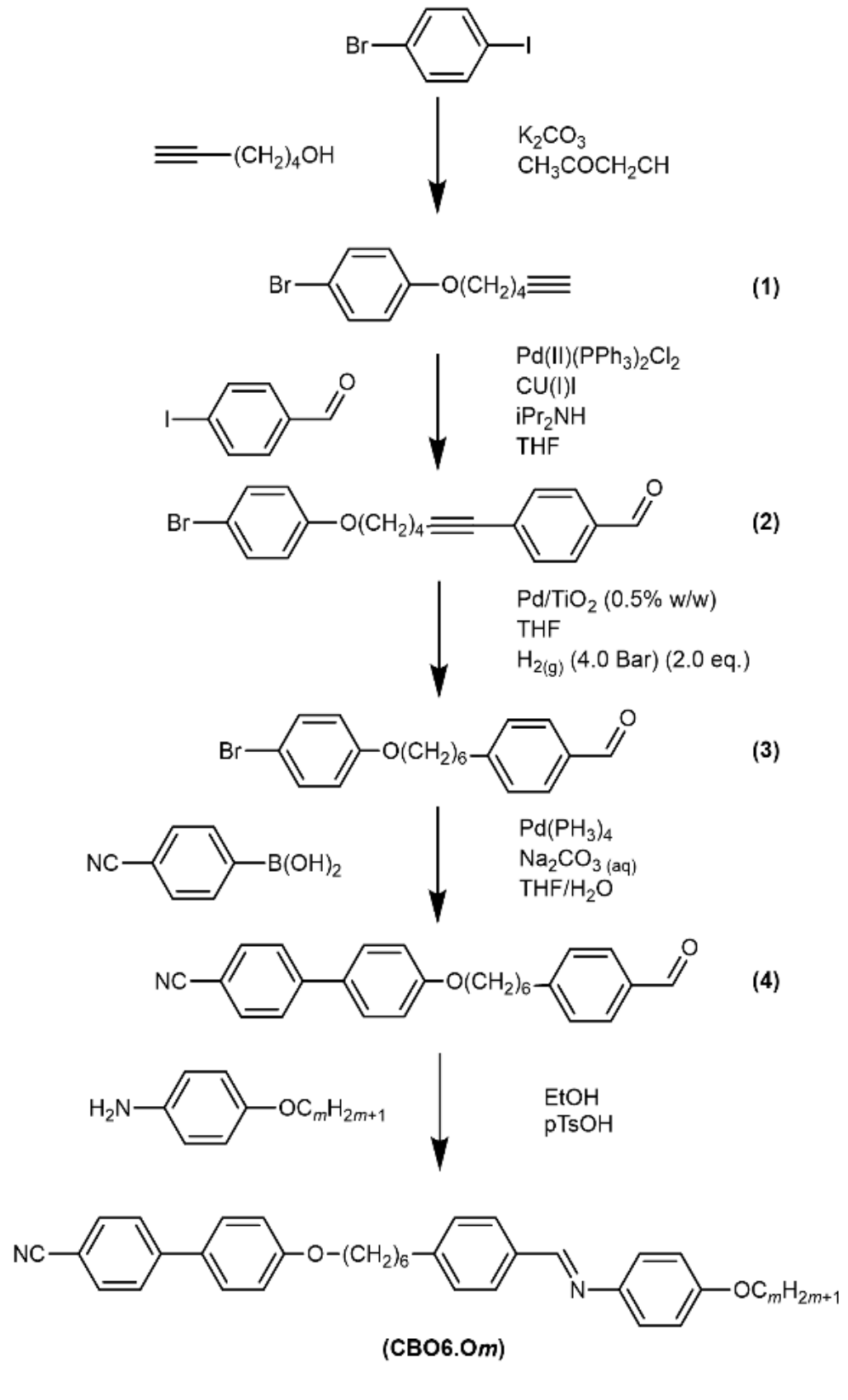

**Scheme 1**. Synthesis of the CBO6.O*m* series, where *m*= 2-4, 6-10.

## Results and Discussion

A summary of the phase properties of the CBO6.O*m* series is given in **Fig. 2** and in tabulated form within the accompanying ESI (**Table S1**). All members of the CBO6.O*m* series display conventional nematic (N) phase. The shortest (*m*=2) and intermediate (*m*=7, 8, 9) homologues of the CBO6.O*m* series also exhibit $N_{TB}$ phase below the N phase with the temperature range of the $N_{TB}$ phase for *m*=8 being the largest. The $N_{TB}$ phase was identified by observation of either a blocky or rope-like textures when viewed between untreated glass slides and the formation of a stripped texture in thin cells with planar anchoring (HG cells) (**Fig. S1**).

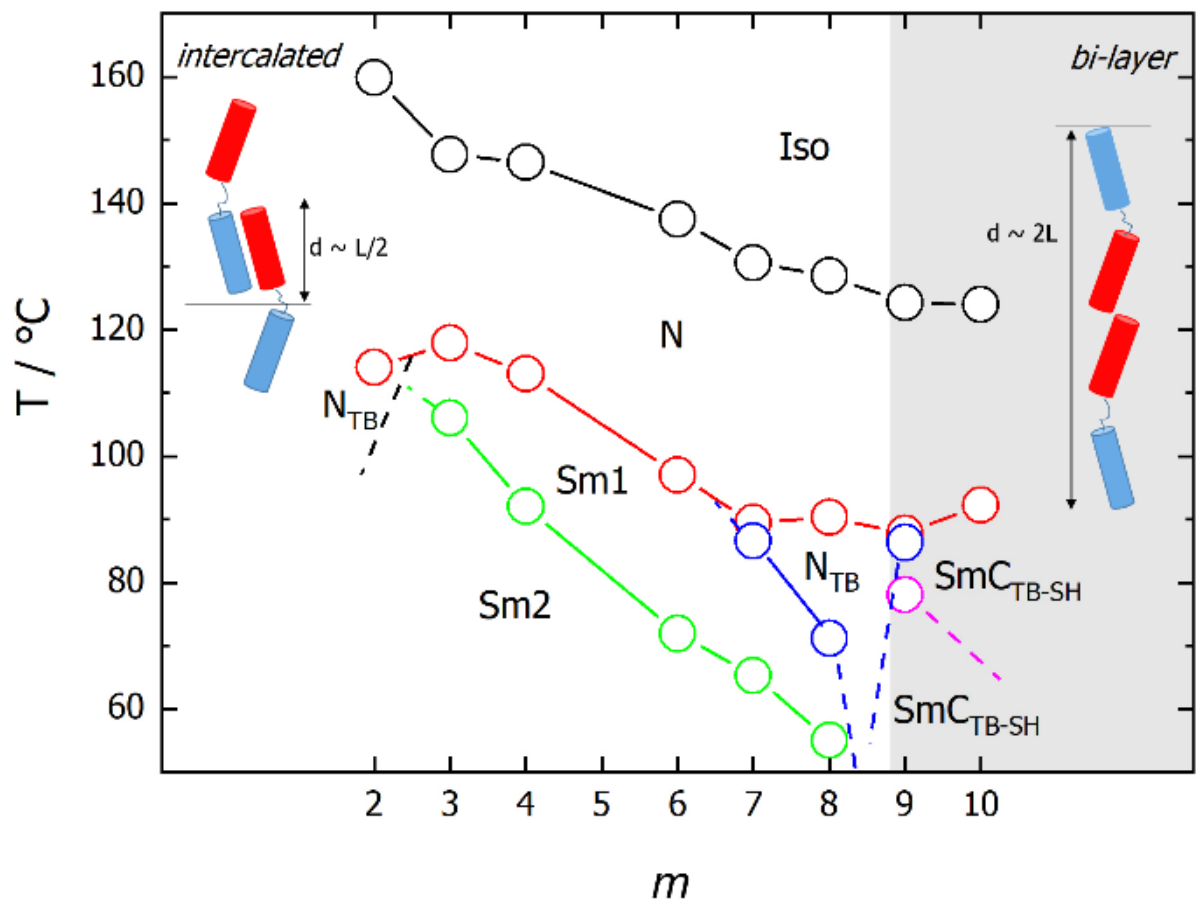

**Figure 2**. Phase diagram for homologue series CBO6.O*m* series. Grey-shaded area marks the homologues exhibiting bi-layer type molecular packing with antiparallel orientation of molecules in consecutive layers. As the insets, sketches of the two different molecular packing schemes: intercalated with d=L/2 and bilayer with d=2L.

The N-$N_{TB}$ phase transition was also evident in the respective DSC thermograms, appearing as weakly first order in nature (**Fig. S2**). The assignment of nematic phases was further confirmed by X-ray diffraction studies (only diffused signals characteristic for liquid like order were registered) and the temperature dependence of optical birefringence (Δn). Specifically, the birefringence follows on cooling in nematic phase the power-law increase, while in the $N_{TB}$ phase it decreases **(Fig. 3**) due to the tilting of the molecules with respect to the helix axis, and azimuthal averaging of molecular positions along the helix.[17]

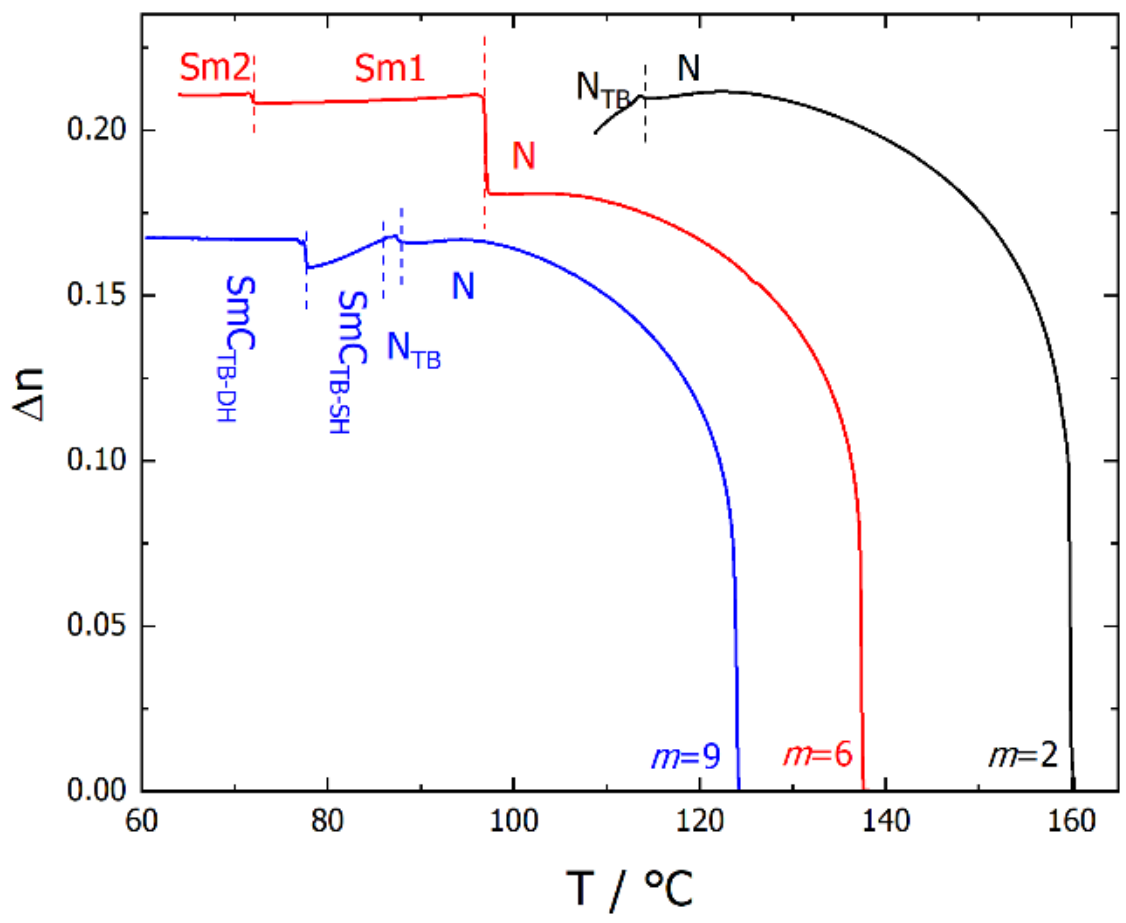

**Figure 3**. Optical birefringence, $\Delta n$, measured with red light ($\lambda$= 633 nm) vs. temperature, for chosen homologues of CBO6.O*m* series.

Further to the nematic phases, a various lamellar LC phases are also formed by homologues of the CBO6.O*m* series - the structures of which are quite complex. For short homologues with *m*=3-8, below N (or $N_{TB}$) phase, a smectic phase was observed, Sm1, which was characterised by a 'dense' schlieren texture observed in thin cells with homeotropic (HT) anchoring condition (**Fig. 4a**).

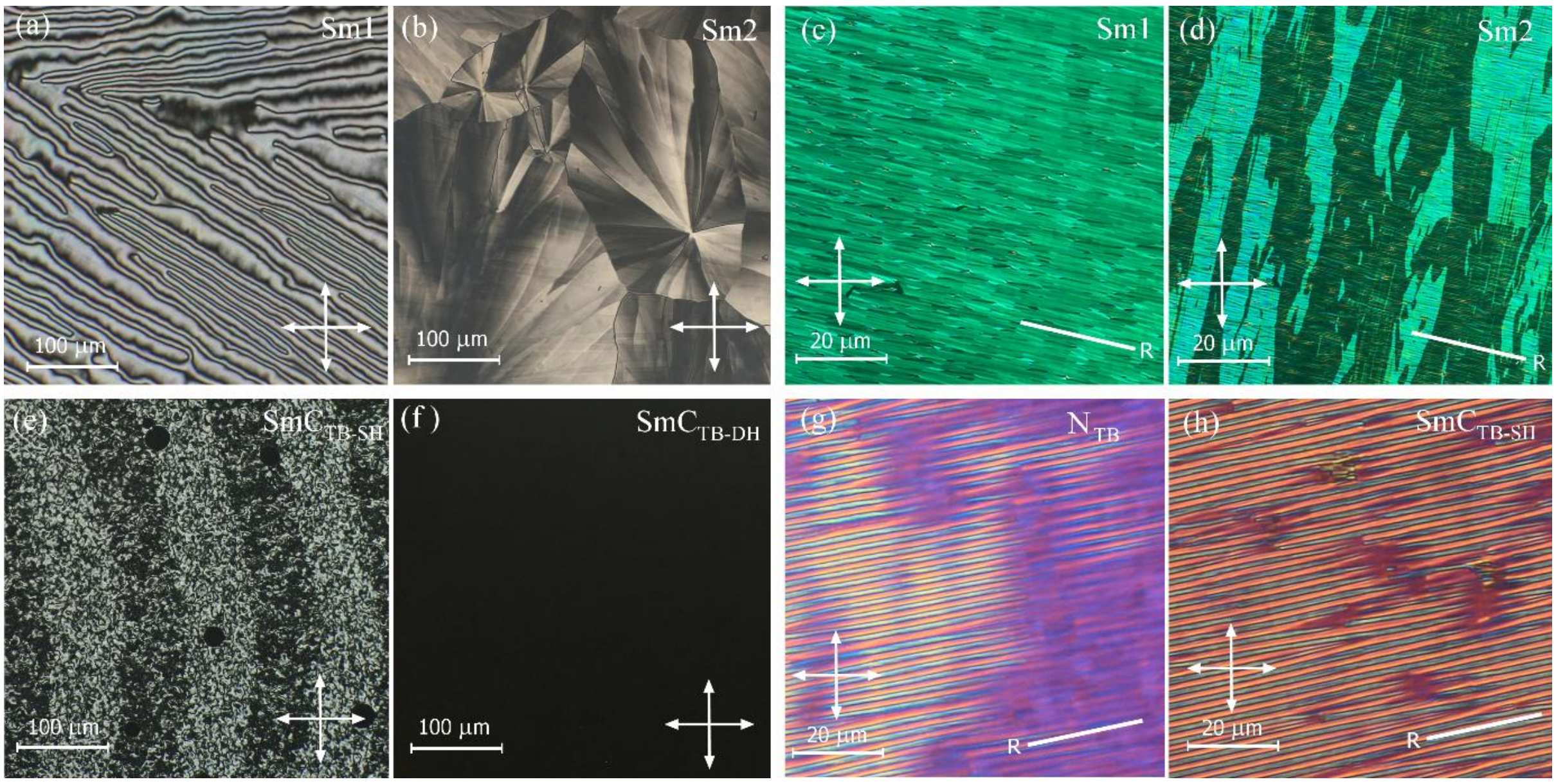

**Fig. 4**: Textures of Sm1 and Sm2 phases of homologue *m*=7 observed between crossed polarizers in **(a, b)** 5-µm-thick cell with homeotropic anchoring (HT cell), and **(c, d)** 3-µm-thick cell with planar anchoring (HG cell). **(e, f)** Textures of $SmC_{TB-SH}$ and $SmC_{TB-DH}$ phases of homologue *m*=9 observed in wedge cell with homeotropic anchoring. **(g, h)** Textures of $N_{TB}$ and $SmC_{TB-SH}$ phases of homologue *m*=9 observed in 3-µm-thick HG cell. Arrows mark polarizers direction and R stands for rubbing direction.

In thin cells with planar anchoring smooth fan textures form (**Fig. 4c**), with the light extinction direction along the rubbing (i.e. the smectic layer normal). On further cooling of the sample a second smectic phase, Sm2 is found, which gives 'plate-like' domains in thin HT cell (**Fig. 4b**). In HG cells the uniform texture of Sm1 transforms into texture with domains for which light extinction direction is inclined from rubbing direction, confirming that the molecules tilt with respect to the layer normal (**Fig. 4d**). The tilt deduced from optical measurements is ~10-15 degrees. The x-ray diffraction (XRD) studies confirmed that for all homologes with $m \leq 8$, the smectic phases have intercalated structure [18], with layer spacing, *d*, equal approximately half the molecular length, *L* (**Fig. 5**). The XRD studies identified the Sm1 phase as lamellar phase with liquid-like order of molecules in the layers and revealed soft-crystalline character of the Sm2 phase, with some degree of positional ordering inside the layers (**Fig. S3**). The same local intercalation is shared by the proceeding nematic phases of these homologues. Summarizing: short homologues of the CBO6.O*m* series tend to form intercalated structures in which the smectic layers are formed by regions of mixed mesogenic cores divided by regions of alkyl chains made of internal and terminal alkyl chains, notably the largest thermal stability of intercalated smectic is found for homologues *m*=3 and 4, for which the length of terminal chain fits half the length of the internal chain, ensuring most efficient molecular packing.

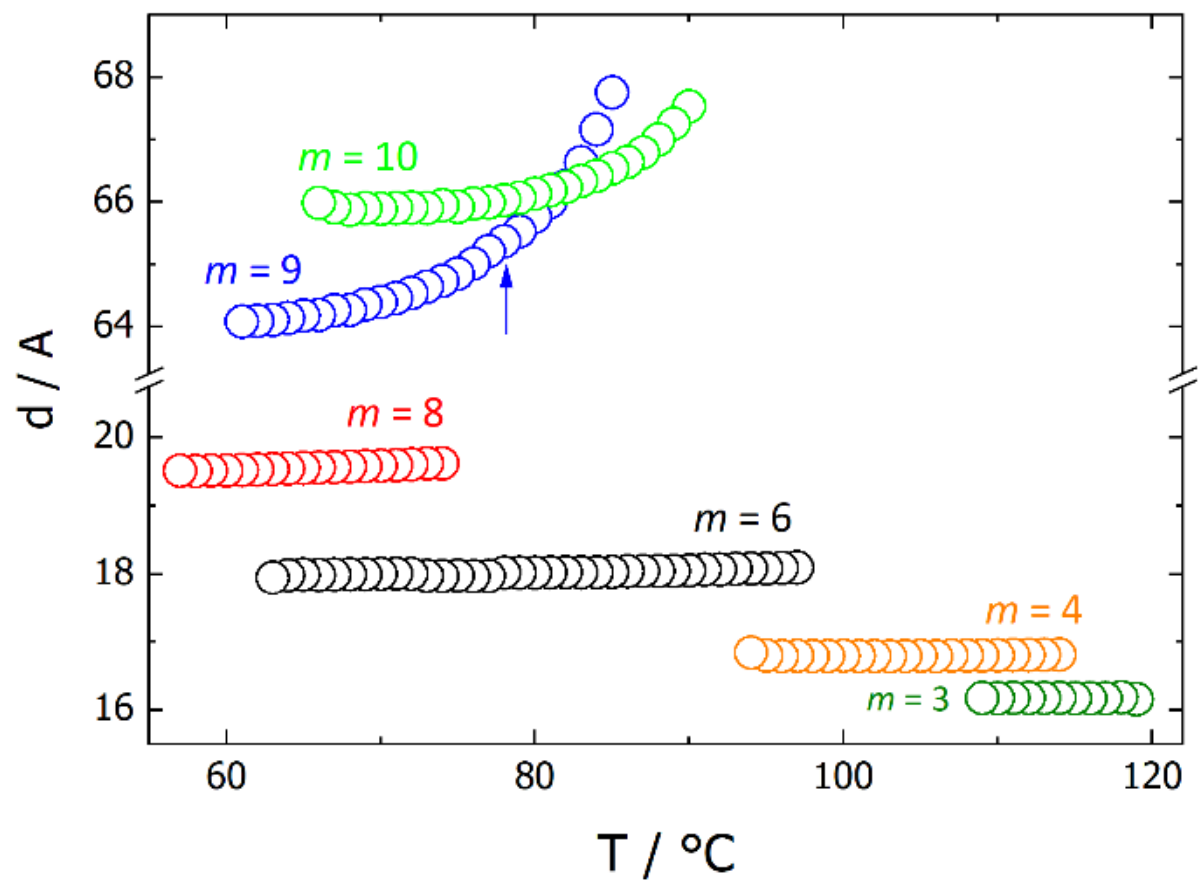

**Figure 5**. Layer spacing in smectic phases vs. temperature for chosen homologues, short homologues *m*= 3-8 form intercalated structure (*d*=L/2) while longer homologes (*m*=9,10) form bilayer structure (*d*=2L). Arrow indicate $SmC_{TB\text{-}SH}$ - $SmC_{TB\text{-}DH}$ phase transition for homologue *m*= 9.

The two longest studied homologues, *m*=9 and *m*=10, behave very differently from their shorter counterparts, forming lamellar phases with layer spacing equal twice the molecular length (**Fig. 5**). This indicates broken up-down symmetry of the molecules in the neighbouring layers.

Apparently, for longer homologues, interactions between like mesogenic cores appear to be favoured. The smectic phase, found directly below the $N_{TB}$ phase for homologue *m*=9 and below the N phase for *m*=10, exhibits birefringent, schlieren-type texture when observed in thin HT cells (**Fig. 4e**), but the texture is distinctly different from those found for the shorter homologues. In wedge cells with homeotropic anchoring, the schlieren texture is covered by strips of different birefringence **(Fig. 4e).** Similar observations have been reported previously for the $SmC_{TB-SH}$ phase, a tilted phase with a short helix with a distorted clock-like type structure. [9] As expected, studies of the temperature dependence of Δn for homologue *m*=9 **(Fig. 2)** show a small, step-like increase in birefringence at N-$N_{TB}$ phase transition, due to increase of orientational order of molecules, followed by its smooth decrease due to the helix formation in $N_{TB}$ and $SmC_{TB-SH}$ phases. [17] This suggests that the tilt angle changes at the $N_{TB}$ -$SmC_{TB-SH}$ phase transition is continuous. In $SmC_{TB-SH}$ phase, on further cooling, Δn values monotonically decrease, indicating growing tilt angle. At the transition to lower-temperature smectic phase, the homeotropic texture in HT cells is restored, pointing to optically uniaxial character of this phase **(Fig. 4f)** and the transition is accompanied by a small jump in Δn (**Fig. 2**) measured in HG cell. The phase transition between smectic phases is not observed in XRD studies, as no change in layer spacing is detected (**Fig. 5)**, and both phases show liquid-like molecular order within the layers (**Fig. S3**). This implies that the transition occurs only by an averaging of azimuthal molecular positions on the tilt cone. The above observations together indicate that the phase below $SmC_{TB-SH}$ is the $SmC_{TB-DH}$ phase, [9] where an additional helical modulation is imposed on the basic 4-layer helical structure **(Fig. 6**). The assumed phase sequence: $N_{TB}$ - $SmC_{TB-SH}$ - $SmC_{TB-DH}$ was confirmed by resonant X-ray scattering measurements (RSoXS, with incident beam energy tuned to the absorption edge of the carbon atom), a technique sensitive to the modulations of molecular orientation.[19] For homologue *m*=9, the resonant diffraction signal in the $N_{TB}$ phase indicates a helix of ~15 nm, in the $SmC_{TB-SH}$ phase the resonant diffraction signal corresponds to the distance of 4 molecular layers ($q_4$=$q_0$/4, where $q_0$ is wavevector of single molecular layer and in the $SmC_{TB-DH}$ phase this signal splits symmetrically into two, positioned at $q=q_4 \pm q_m$, where $q_m$ is the wavevector of the additional helical modulation superimposed on the 4-layer helix. The additional helix winds as the temperature decreases, from ~100 nm at the $SmC_{TB-SH}$ - $SmC_{TB-DH}$ phase transition, to ~40 nm before sample crystallization, ~4 K below. The slight asymmetry in the split signal intensities indicates a distorted clock-like structure for the 4-layer basic unit [9,15].

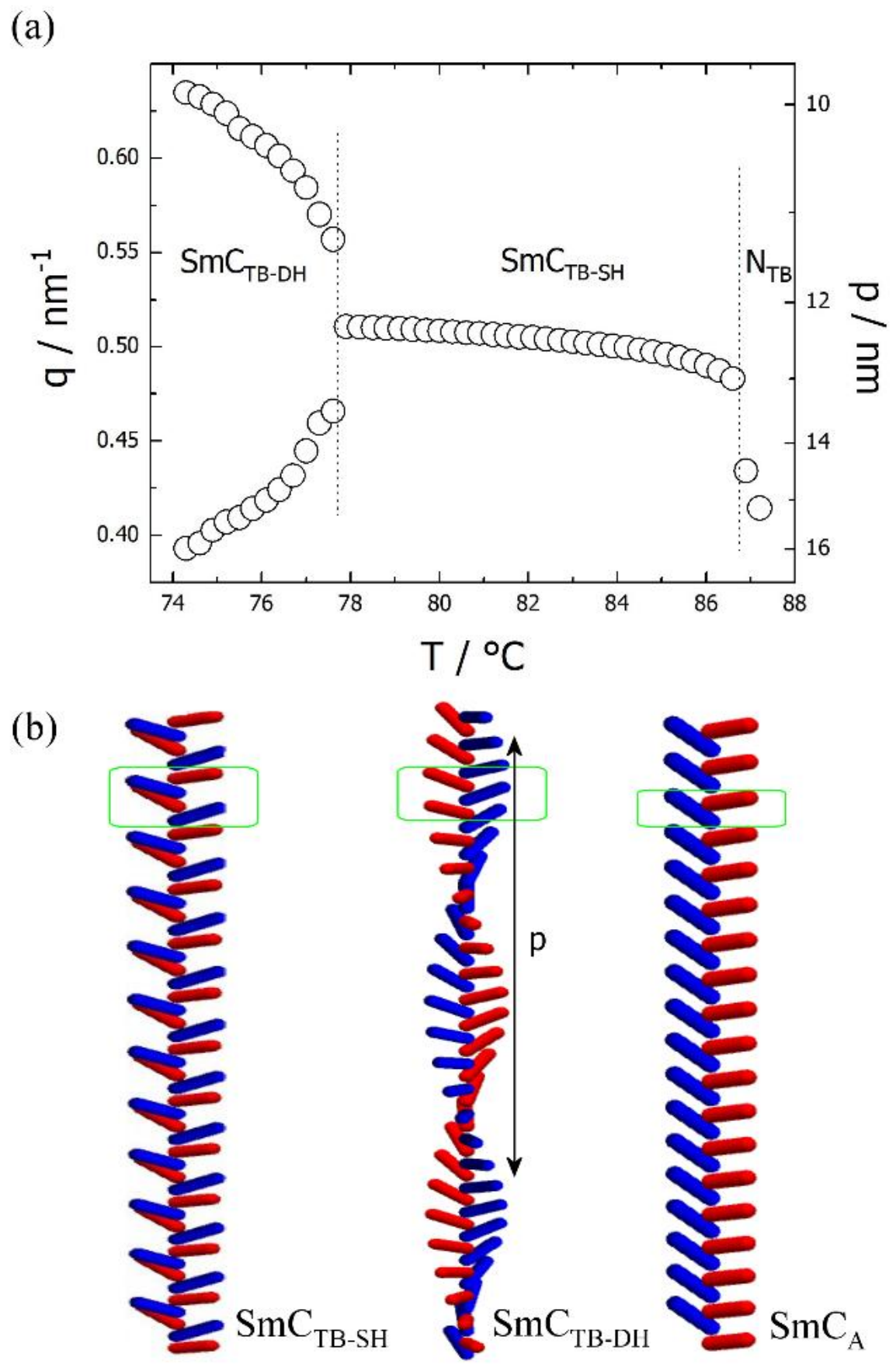

**Fig. 6. (a)** Wavevector, q, and corresponding periodicity, p, determined by resonant x-ray diffraction studies for compound m=9. **(b)** model of $SmC_{TB-SH}$ with 4-layer distorted clock like unit (green box) and $SmC_{TB-DH}$ phase in which additional helical modulation (helical pitch p) was superimposed on the basic 4-layer periodicity. The $SmC_{TB-DH}$ is intermediate state between anticlinic $SmC_A$ and $SmC_{TB-SH}$ with four layers structures. For simplicity the dimers in schematic drawing were represented by their long axis (the line connecting molecular ends) and the tilt angle value was exaggerated to make the modulated structure more easily seen. Only azimuthal angle distribution was shown.

Summarizing, similarly as previously studied dimeric mesogens the CBO6.m homologue series shows two distinctive packing motifs: short homologues $m \leq 8$ form intercalated lamellar structures, while long homologues, $m>8$, form bi-layer-type smectics. As a result of these competing tendencies, intermediate homologues ($m$=7-9) "revisit" the nematic phase (N and $N_{TB}$). Miscibility tests, using homologues with $m$=6 and $m$=9, within a contact cell showed that the Sm1 and $SmC_{TB-SH}$ phases are not miscible, with only the N and $N_{TB}$ phases being observed in the contact area (**Fig. S4**).

As a final point, let us compare the transitional properties of the CBO6.O*m* materials presented here with their structural isomers, the previously reported CB6O.O*m* series [13,14]. Both series share similar transitional properties, with the $N_{TB}$, $SmC_{TB\text{-}SH}$, and $SmC_{TB\text{-}DH}$ phases, despite the difference in molecular shape afforded by the change of the spacer. The CBO6.O*m* molecules adopt a more ‘v-shaped’ conformation whereas the CB6O.O*m* analogues adopt more of a ‘hockey-stick’ shape (**Fig. 7a**).

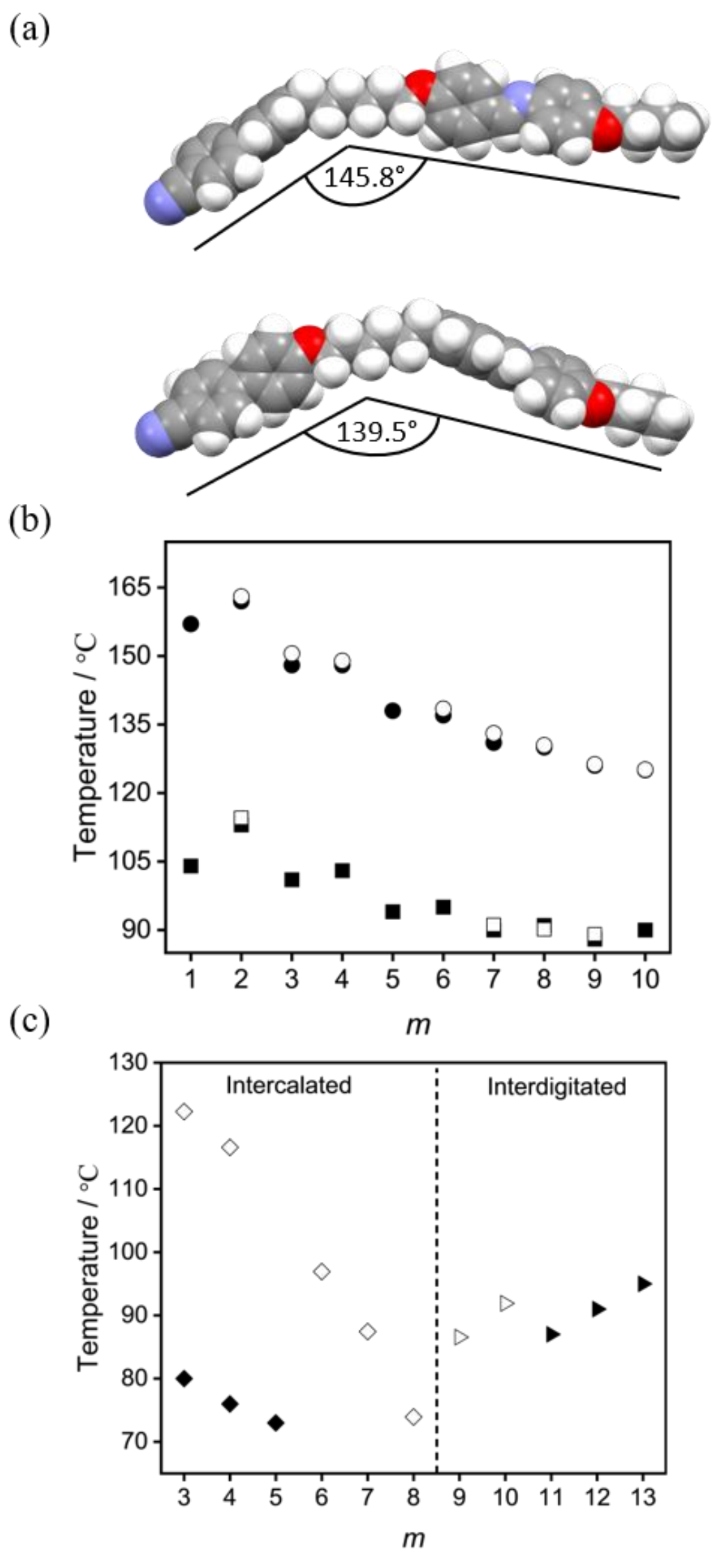

**Fig. 7. (a)** Energy minimised space filling models calculated at the 6-31G(d) level of theory comparing the molecular shapes of CB6O.O4 (top) and CBO6.O4 (bottom). Comparison of the values of **(b)** $T_{N\text{-}I}$ [circle] and $T_{NTB}$ [square]; **(c)** $T_{Sm1}$ [diamond] and $T_{SmCTB\text{-}SH}$ [triangle] for the CB6O.O*m* and CBO6.O*m* series. The CB6O.O*m* are represented by the solid symbols and the CBO6.O*m* by the open symbols.

It therefore might be expected that the change in shape would be reflected in the thermal properties of these materials, particularly in the $N_{TB}$-N transition temperatures ($T_{NTB-N}$) - especially given the view that the formation of the $N_{TB}$ phase is largely driven by molecular shape. [5,20] Beyond our expectations, for a given terminal chain length (*m*), there is no significant differences between the values of $T_{NTB-N}$ (**Fig. 7b**), and the same observation is seen for the nematic-isotropic transition temperatures ($T_{N-I}$). However, there are significant differences between CB6O.Om and CB6O.Om series seen in thermal stability of smectic phases (**Fig. 7c**). Although the materials presented here do display a greater number of smectic phases at shorter terminal chain lengths, the associated transition temperatures are ~40 K higher for a given value of *m*. Interestingly, a similar packing dependence on length of the terminal chain is seen between the two series. Both series display smectic phases with intercalated structures for short homologues, however, in the CB6O.O*m* series smectic polymorphism is lost at *m*=6 until the emergence of bilayer-type smectic phases at *m*=11. When the spacer is reversed in CBO6.O*m* materials, smectic phases are observed continuously from *m*=3 with the switch from an intercalated to bilayer arrangement occurring at *m*>8. Seemingly, the reversal of the spacer in the CBO6.Om series promotes and stabilises the formation of lamellar structure and is likely a product of the enhanced polarizability of the cyanobiphenyl (CB) unit afforded by the conjugation of the *p*-orbitals of the oxygen atom into the aromatic π system.

## Conclusions

The flexible CBO6.O*m* dimers have two distinctly different molecular packing schemes: short homologues (*m*<8) form intercalated, while longer homologues (*m*>8) - bilayer structures, and the frustration arising from this leads to a nematic gap for intermediate homologue. Similar behaviour has also been observed for other series of linear and non-linear dimers. [18,21] The dimers studied here are bent-shaped, as the mesogenic cores are connected by an odd number atom spacer; such dimers spontaneously form helical structures, which is not unique to the nematic $N_{TB}$ phase; helices have also been encountered in smectic phases. The heliconical structure may be simple analogue to that of the $N_{TB}$ phase ($SmC_{TB-SH}$) but can also be more complex with additional helix ($SmC_{TB-DH}$). The tendency to form short helical structures or more complex double helices seems to be a universal behaviour of bent dimers. One might ask: what is the driving force that transforms the system into such a complex double-helical structure? The double-helical structure, where a longer helix is superimposed on top of a short four-layer helix, is likely the transient state for the system evolving from the four-layer to a simple anticlinic structure [9]. In chiral rod-like materials, such change is realized by a first-order transition (phase transition between $SmC_A$ and $SmC_{Fi2}$ [22]), while in bent dimers, it is achieved through the formation of the intermediate double-helical structure. Finally, we compare the transitional properties of the materials presented here to their previously reported structural isomers, the CB6O.O*m* series. Both series exhibit similar changes of the packing motifs, intercalated vs. bilayer, with increasing terminal chain length, but the CBO6.O*m* series show considerably higher transition temperatures due to the oxygen atom in the spacer contributing to the polarizability of the CB unit. Despite these differences in the smectic transitions, there are no differences in the nematic transition temperatures (both N and $N_{TB}$) even though molecules of both series have different shapes.

## Acknowledgements

The work was financed by the National Centre for Research and Development (Poland) through the project EIG CONCERT-JAPAN/9/89/FerroFluid/2023 (MM). The beamline 11.0.1.2 at the Advanced Light Source at the Lawrence Berkeley National Laboratory is supported by the Director of the Office of Science, Office of Basic Energy Sciences, of the U.S. Department of Energy under Contract No. DE-AC02-05CH11231.

*Supporting Information for*

# Frustration in system with intercalated and bilayer smectic structure

Calum J. Gibb, Claudio Flores Fernandez, Magdalena M. Majewska, John M. D. Storey, Corrie T. Imrie, Ewa Gorecka, Damian Pociecha

# 1 Methods

## 1.1 *Chemical Synthesis*

Chemicals were purchased from commercial suppliers (Fluorochem, Merck) and used as received. Solvents were purchased from Merck and used without further purification. Reactions were performed in standard laboratory glassware at ambient temperature and atmosphere and were monitored by TLC with an appropriate eluent and visualised with 254 nm light. Chromatographic purification was performed using a Biotage Selekt MPLC using DCM - Pet. Ether (40-60) gradients as the mobile phase, with detection made in the 200-800 nm range.

## 1.2 *Chemical Characterisation*

Chemical materials were characterised by NMR spectroscopy using a Bruker Avance III HD 9.4 T NMR spectrometer operating at 400 MHz, 101 MHz (1H, and 13C{1H}, respectively). All spectra were recorded in $CDCl_3$, and the chemical shifts are given in ppm. Purity analysis was performed by mass spectrometry using a Waters XEVO G2QT of Mass Spectrometer.

## 1.3 *Polarized light microscopy (POM) and birefringence measurements*

Characteristic optical textures of various LC phases were observed with a Zeiss AxioImager A2m microscope. Samples were prepared either between untreated glass plates or in commercially available cells (provided by Warsaw Military University of Technology, WAT), having planar (HG cells) or homeotropic (HT cells) anchoring condition, cells were of 3 – 5 micron thickness. Temperature of the samples was controlled with a Linkam LTS-350 heating stage. For determination of optical birefringence a setup based on a photoelastic modulator (PEM-90, Hinds) was used; with a halogen lamp (Hamamatsu LC8) equipped with narrow bandpass filters ($\lambda$=633 nm) as a light source. Intensity of the transmitted light was monitored by a photodiode (FLC Electronics PIN-20) and the signal was deconvoluted with a lock-in amplifier (EG&G 7265) into 1f and 2f components (where f=50 kHz is the modulation frequency of PEM) to yield a retardation induced by the sample, $Ret = 2\pi\frac{\Delta n \cdot d}{\lambda}$. Samples were prepared in 3-μm-thick HG cells.

## *1.4 Differential scanning calorimetry (DSC)*

Characterisation of the translational properties of the CBO6.O*m* series were determined by differential scanning calorimetry using a Mettler Toledo DSC3 differential scanning calorimeter equipped with a TSO 801RO autosampler and calibrated using an indium standard. Samples were measured under a nitrogen atmosphere with 20 °C $min^{-1}$ heating and cooling rates. The transition temperatures and entropy values reported are averages obtained for duplicate runs and normally extracted from heating traces.

## *1.5 X-ray diffraction*

The X-ray diffraction patterns in broad angle range were registered with a Bruker D8 GADDS system (ceramic tube with copper anode, parabolic mirror monochromator, 0.5-mm point beam collimator, area detector VANTEC2000). Samples were prepared as droplets on a heated surface, their temperature was controlled with a modified Linkam heating stage. Temperature dependence of smectic layer thickness was determined from small angle X-ray diffraction experiments performed with a Bruker Nanostar system (microfocus-type tube with copper anode and dedicated Montel optics, area detector VANTEC2000). Powder-like samples were prepared in thin-walled glass capillaries.

The resonant x-ray diffraction experiments were performed at the Advanced Light Source of Lawrence Berkeley National Laboratory. The soft x-ray scattering beam line (11.0.1.2) was used with the energy of incident beam tuned to the K-edge of carbon absorption band (~283 eV). Samples were prepared between two 100-nm-thick Si3N4 membranes (Norcada). The diffraction patterns were recorded with the cooled CCD detector (Princeton PI-MTE, pixel size of 27 μm).

## *1.6 Atomic force microscopy (AFM)*

The ***AFM images*** were registered with the Bruker Dimension Icon microscope, working in the tapping mode at the liquid crystalline-air surface. Cantilevers with a low spring constant $0.4\ \mathrm{N/m}$ were used, the resonant frequency was in a range of $70 - 80$ kHz, a typical scan frequency was 1 Hz. Samples for the AFM imaging were prepared on quarts substrates.

### *1.7 Molecular modelling*

Quantum mechanical density functional theory (DFT) calculations at the B3LYP 6-31G(d) level of theory have been performed for all members of the CBO6.O*m* series using Gaussian09 software. Optimised geometries, molecular dipole moments and electrostatic potential surfaces were calculated, and visualised using GaussView 5 Space-filling and ball-and-stick models were produced from the optimised geometry using Mercury 4.0.

## 2 Supplementary Data

**Table S1.** Transition temperatures (°C) and associated scaled entropy changes, in brackets, for the CBO6.O*m* series

| *m* | K / °C | *Sm1-Sm2* / °C | *Sm2*-N[a] or $N_{TB}$[b] / °C | $SmC_{TB-DH}$-$SmC_{TB-SH}$ / °C | $SmC_{TB-SH}$-N[a] or $N_{TB}$[b] / °C | $N_{TB}$-N / °C | N-I / °C |
|---|---|---|---|---|---|---|---|
| *2* | 117.6 (6.26) | | | | | [114.6] (0.01) | 163.0 (0.21) |
| *3* | 113.9 (3.33) | 133.9 (3.33) | 122.3 (2.48)[a] | | | | 150.5 (0.17) |
| *4* | 117. 3 (7.61) | [97.0] (2.10) | [116.6] (2.18)[a] | | | | 148.9 (0.23) |
| | | | | | | | |
| *6* | 107.7 (6.46) | [71.8] (0.97) | [96.9] (1.44)[b] | | | | 138.4 (0.17) |
| *7* | 107.0 (7.70) | [65.3] (1.03) | [87.5] (1.12)[b] | | | [91.1] (0.01) | 133.1 (0.15) |
| *8* | 94.4 (7.47) | [56.0] (0.97) | [73.9] (0.73)[b] | | | [90.1] (0.002) | 130.4 (0.18) |
| *9* | 85.5 (13.2) | | | [81.0][†] (≈0) | 86.6 (0.01)[a] | 89.1 (0.001) | 125.9 (0.16) |
| *10* | 88.7 (13.4) | | | | [91.9] (0.10)[b] | | 125.1 (0.17) |

*[†] Microscope observation, [] monotropic transition.*

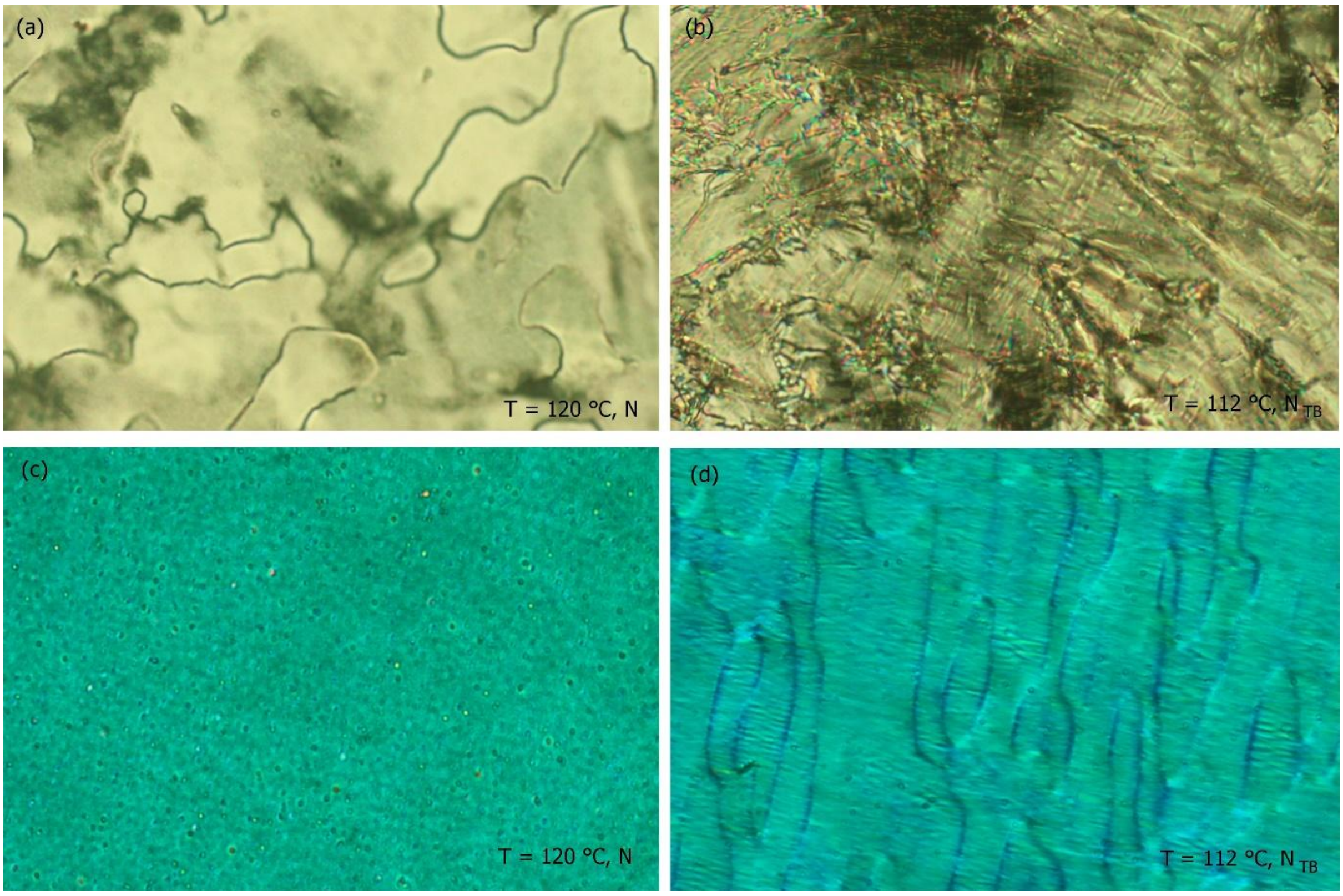

**Figure S1.** Optical texture of **(a, c)** N and **(b, d)** $N_{TB}$ phases observed for homologue *m*= 2 between untreated glass plates **(a, b)** or in 3-μm-thick cell with planar anchoring condition.

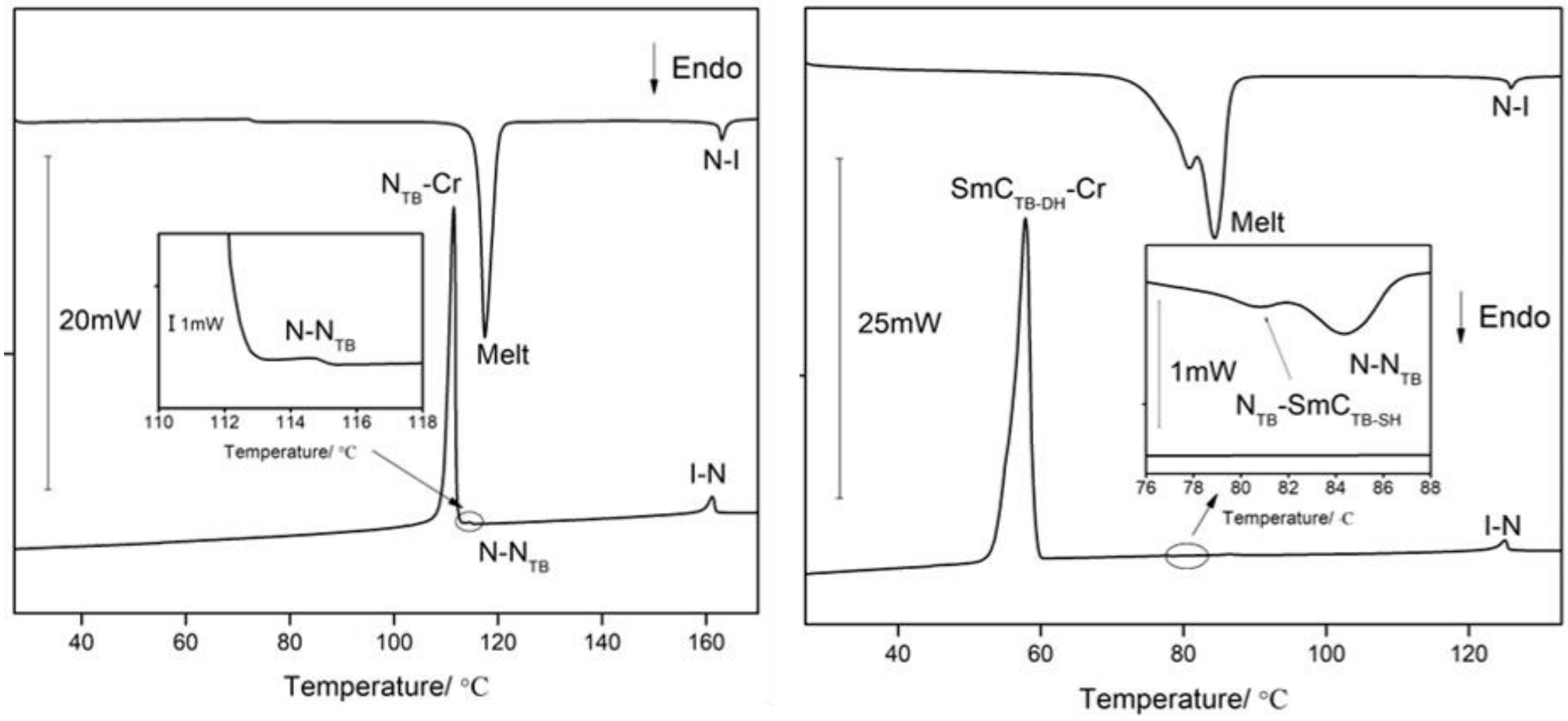

**Figure S2.** Representative DSC thermograms for CBO6.O2 [left] and CBO6.O9 [right].

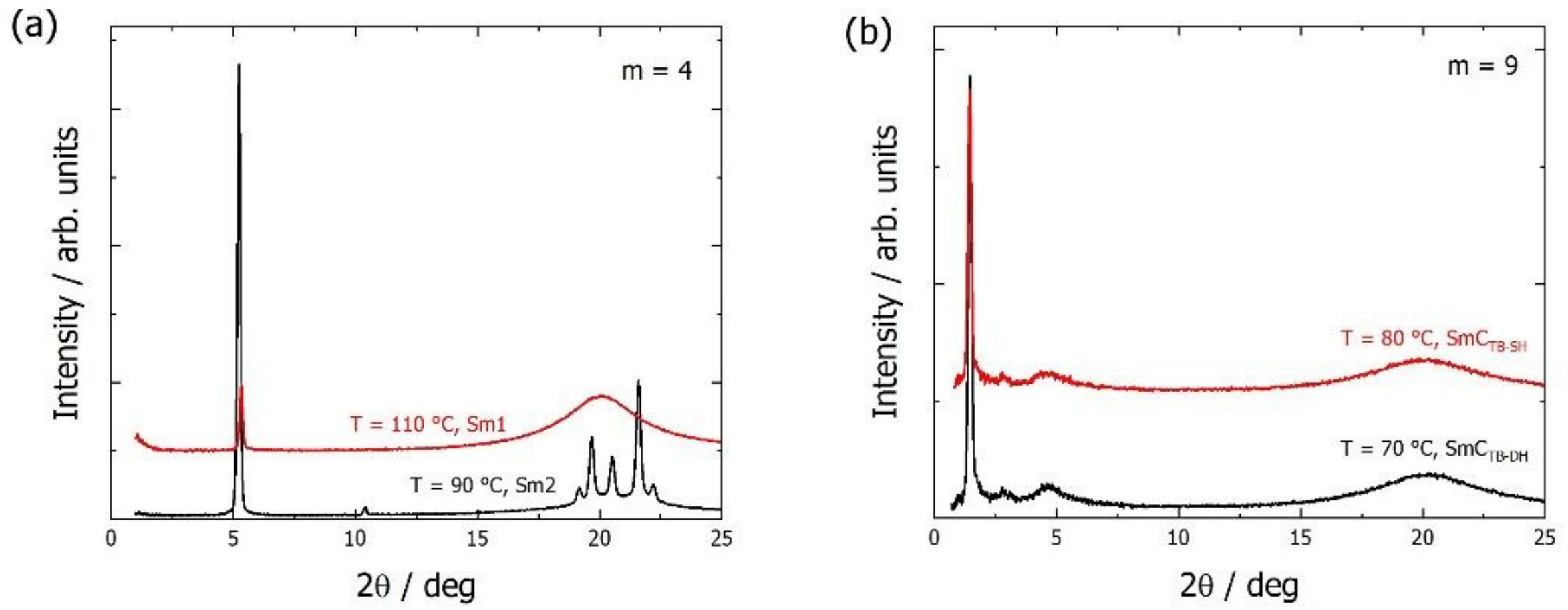

**Figure S3.** X-ray diffraction patterns for (a) Sm1 and Sm2 for homologue m=4, the main signal detected at small diffraction angles corresponds to half molecular length, d~L/2, and (b) in $SmC_{TB-SH}$ and $SmC_{TB-DH}$ for homologue m=9. Note, that in small angle region apart from the main signal related to the bilayer structure with d~2L also weak, a diffused signal centred at d=L/2 is observed, suggesting some local intercalated correlations of molecules. Diffused signals detected at high angles in Sm1 and both $SmC_{TB}$ phases show the liquid-like correlations inside the smectic layers. Sharp signals in Sm2 phase indicate that phase is soft crystal type.

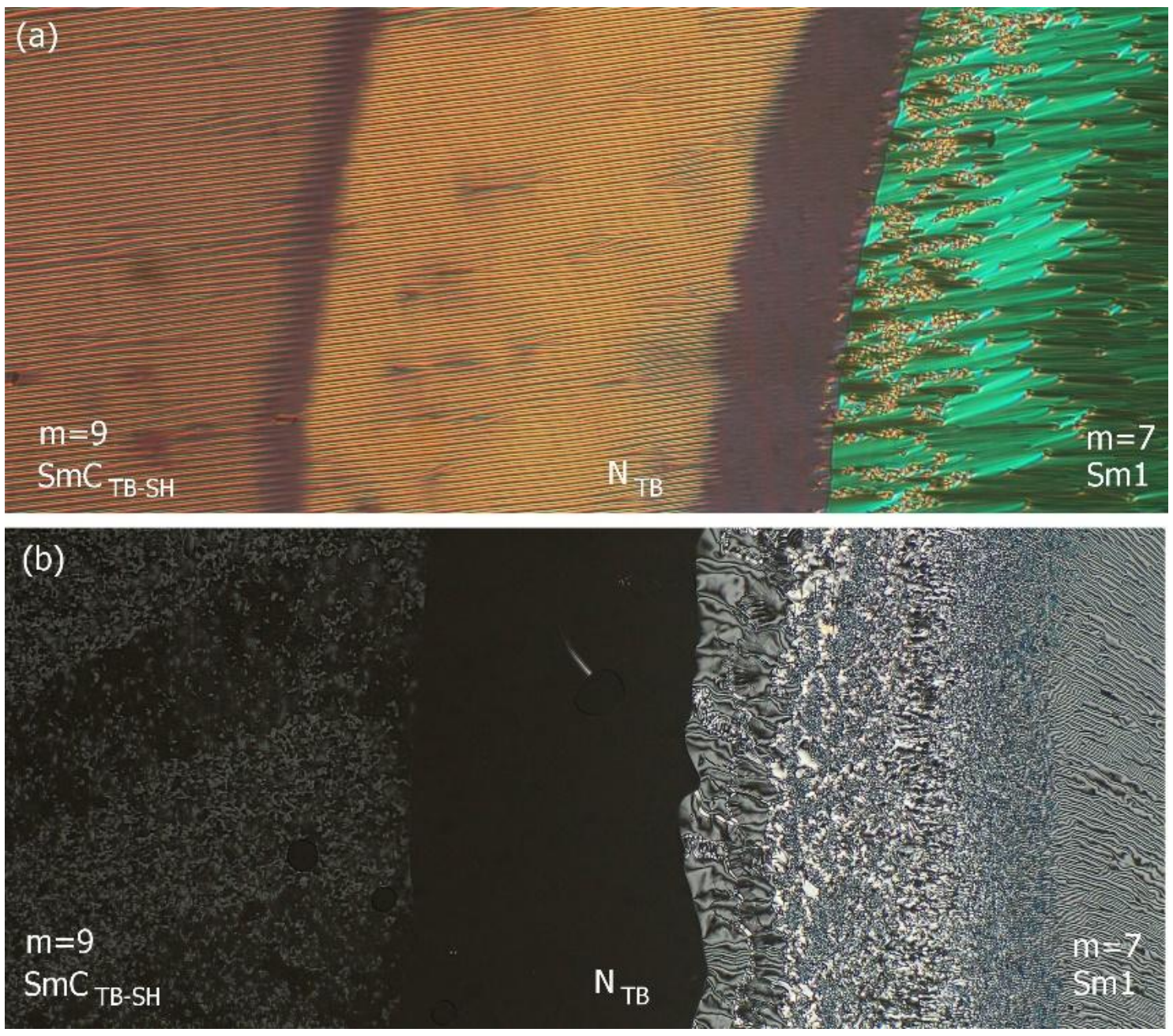

**Figure S4.** Optical texture of contact cell of homologue m=9 (left) and m=7(right) taken at 80 °C, in (a) 3-µm-thick HG cell and (b) 5-µm-thick HT cell. Apparently, bilayer-type $SmC_{TB-SH}$ and intercalated Sm1 are not miscible, in the contact area the $N_{TB}$ phase is formed.

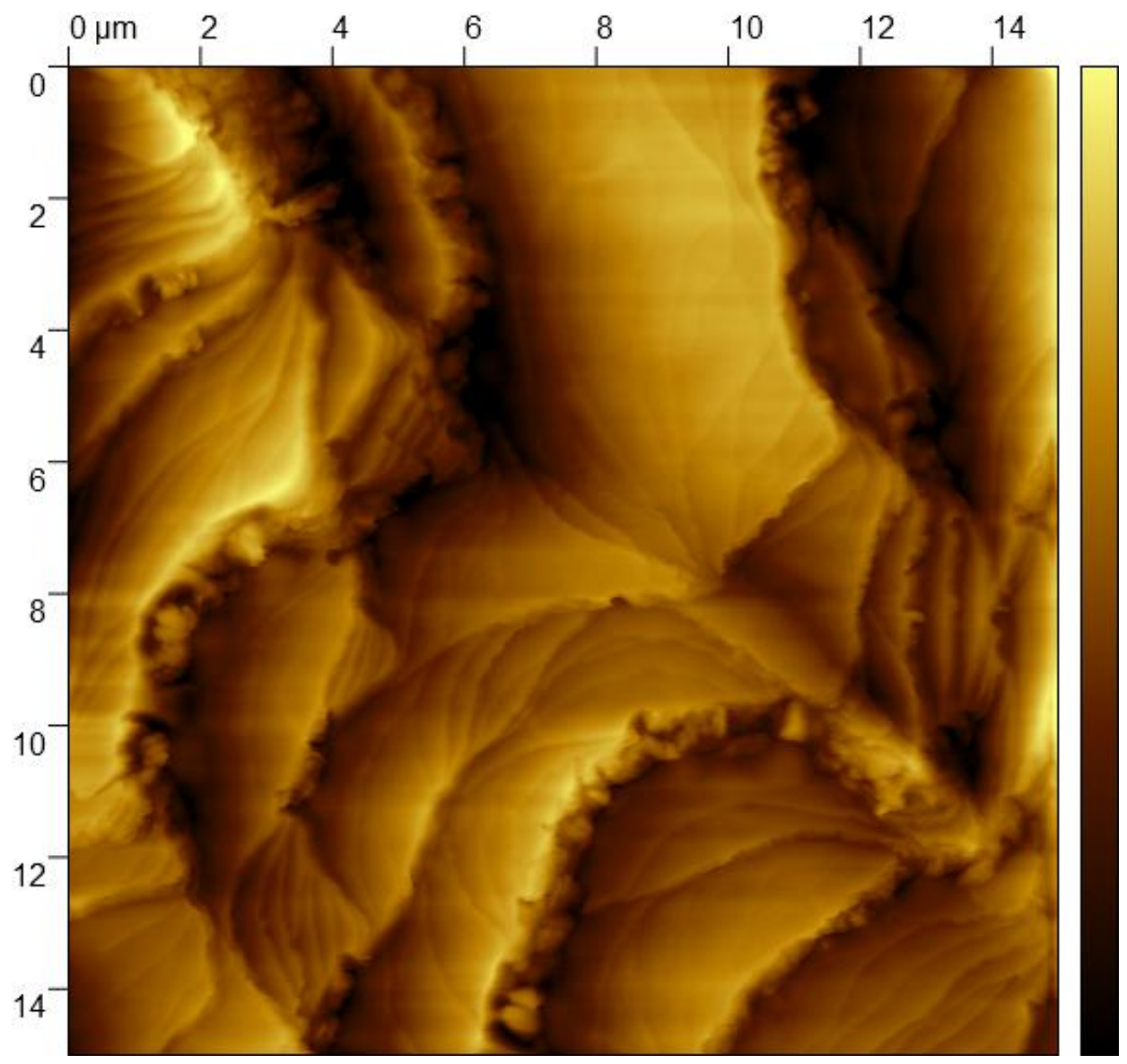

**Figure S5.** Surface topography of Sm2 phase of homologue m=6, determined by AFM method. The sample was placed on the quartz plate and quenched to room temperature.

# 3 Chemical Synthesis

The CBO6.O*m* series was synthesised using the synthetic scheme shown in **Scheme 1** (in main manuscript).

## 3.1 Preparation of 1-bromo-4-(hex-5-yn-1-yloxy)benzene (1)

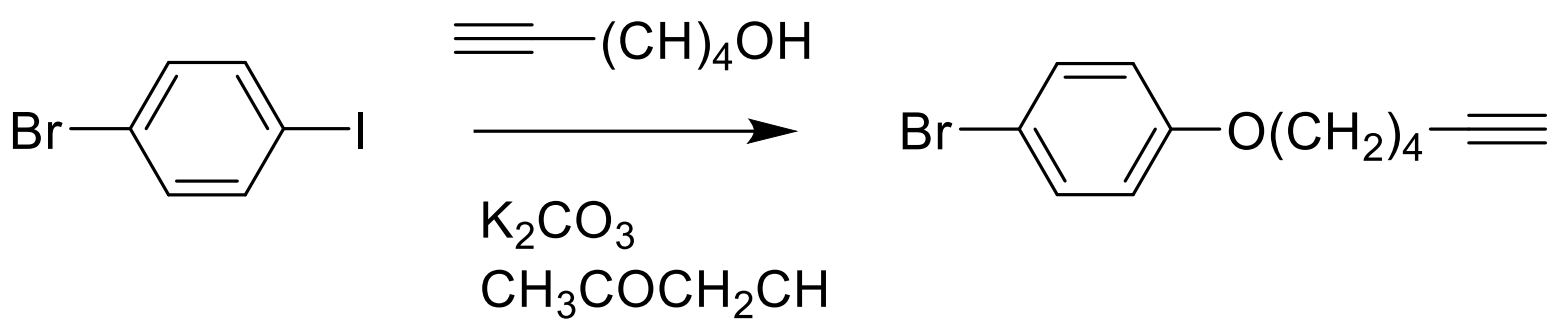

**Scheme S1.** Synthesis of 1-bromo-4-(hex-5-yn-1-yloxy)benzene **(1)**

The synthesis of the 1-bromo-4-(hex-5-yn-1-yloxy)benzene **(1)** is shown in **Scheme S1**. 5-hexyne-1-ol (2.00 g, 2.25 mL, 21.0 mmol, 1eq.) and 4-bromophenol (4.00 g, 23.1 mmol, 1.1eq.) were dissolved in 20mL of butanone. $K_2CO_3$ (5.80 g, 42.0 mmol, 2eq.) was added and the reaction heated to 70 °C for 12 h. The reaction was then cooled to room temperature with the $K_2CO_3$ being removed by filtration and the butanone being removed under reduced pressure. The crude product of I6.1.6 was then dissolved in 20 mL DCM and washed: 2x 10 mL 0.5M

NaOH and 1x 10 mL of brine. The sample was then purified by column chromatography using DCM : Hexane (3:7) as the eluent to give 1 bromo 4 (hex 5-yn-1-yloxy)benzene as a colourless oil which was dried under reduced pressure overnight to give the reported yield.

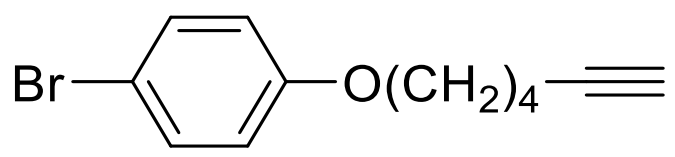

**(1)**

1-bromo-4-(hex-5-yn-1-yloxy)benzene

| | |
|---|---|
| Yield: | (colourless oil) 4.83 g, 90 %. |
| $R_f$ (DCM: Hexanes [3:10]): | 0.47 |
| $^1$H NMR (400 MHz): | 7.36 (dd, *J* = 8.9, 2.0 Hz, 2H, Ar-**H**), 6.77 (dd, *J* = 8.8, 2.0 Hz, 2H, Ar-**H**), 3.95 (t, *J* = 6.2 Hz, 2H, Ar-O-C**H**$_2$-CH$_2$), 2.32 – 2.22 (m, 2H, C**H**$_2$-C≡C-H), 1.97 (t, *J* = 2.7 Hz, 1H, C≡C-**H**), 1.90 (p, *J* = 6.5 Hz, 2H, Ar-CH$_2$-C**H**$_2$-CH$_2$), 1.71 (p, *J* = 7.1 Hz, 2H, CH$_2$-C**H**$_2$-CH$_2$-C≡C-H). |
| $^{13}$C{$^1$H} NMR (101 MHz): | 158.07, 132.19, 116.25, 112.69, 83.97, 68.68, 67.51, 28.16, 24.96, 18.13. |

### 3.2 Preparation of 4-(6-(4'-bromophenoxy)hex-1-yn-1-yl)benzaldehyde (2)

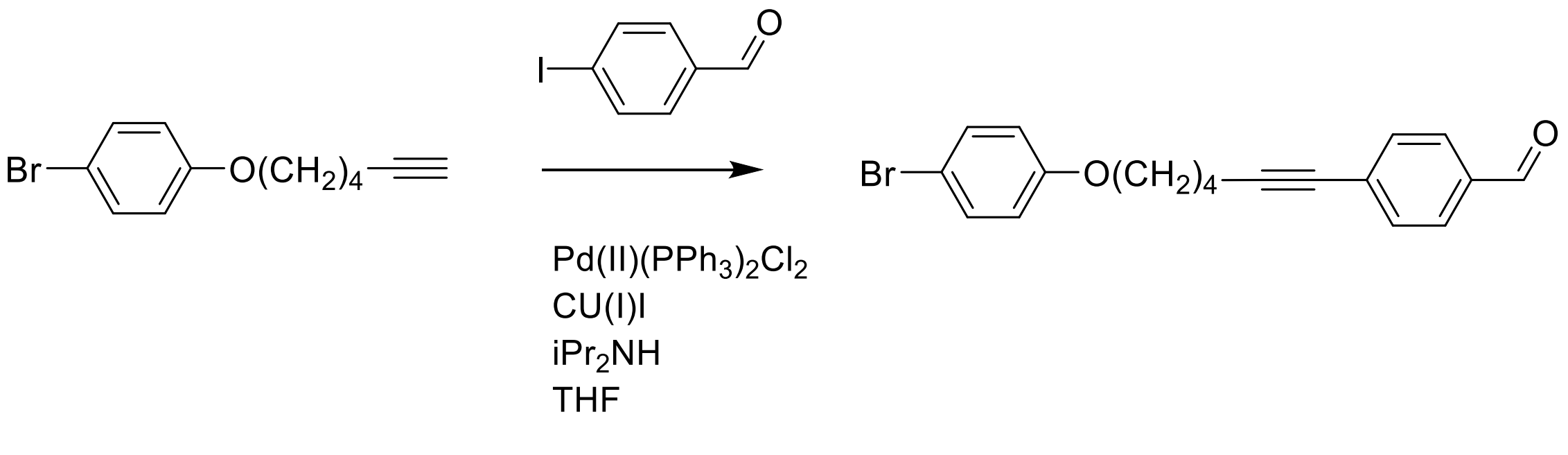

**Scheme S2.** Synthesis of 4-(6-(4'-bromophenoxy)hex-1-yn-1-yl)benzaldehyde (**2**)

The synthesis of the 4-(6-(4'-bromophenoxy)hex-1-yn-1-yl)benzaldehyde (**2**) is shown in **Scheme S2**. (**1**) (4.80 g, 19 mmol, 1.0 eq.) and 4-Iodobenzaldehyde (4.66g, 20mmol, 1.1 eq.) were added to a flame-dried flask and dried by azeotrope with dry toluene. The dried solids were then dissolved in 5 mL of dry THF which was then sparged with $Ar_{(g)}$ for 30 mins. Bis(triphenylphosphine)palladium(II) dichloride (266 mg, 3.8 mmol, 0.02 eq.) and copper(I) iodide (72 mg, 3.8 mmol, 0.02 eq.) were then added followed by the dropwise addition of *n*,*n*-diisopropylethylamine (DIEA) (2.13 g, 3.0 mL, 21 mmol, 1.1 eq.) over 15 mins. The reaction was then heated to 40 °C for 1 h with the completion of the reaction being confirmed by TLC (EtOAc : Pet. Ether [1:9]). The reaction was then cooled to room temperature, filtered and the THF removed under reduced pressure. The crude mixture was then dissolved in 20 mL DCM and washed: 2x 10mL of brine before being purified further by flash chromatography using a Biotage Selekt MPLC using DCM : Pet. Ether [1:9] as the eluent. 4-(6-(4'-bromophenoxy)hex-1-yn-1-yl)benzaldehyde was retrieved as a pale-yellow oil which was dried under reduced pressure overnight to give the reported yield.

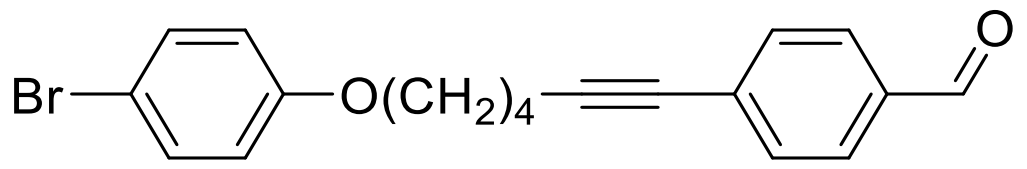

**(2)**

4-(6-(4'-bromophenoxy)hex-1-yn-1-yl)benzaldehyde

Yield: (pail-yellow oil) 5.70 g, 84 %.

$R_f$ (EtoAc: Hexanes [1:9]): 0.35

$^1$H NMR (400 MHz): 7.36 (dd, *J* = 8.9, 2.0 Hz, 2H, Ar-**H**), 6.77 (dd, *J* = 8.8, 2.0 Hz, 2H, Ar-**H**), 3.95 (t, *J* = 6.2 Hz, 2H, Ar-O-C**H**$_2$-CH$_2$), 2.32 – 2.22 (m, 2H, C**H**$_2$-C≡C-H), 1.97 (t, *J* = 2.7 Hz, 1H, C≡C-**H**), 1.90 (p, *J* = 6.5 Hz, 2H, Ar-CH$_2$-C**H**$_2$-CH$_2$), 1.71 (p, *J* = 7.1 Hz, 2H, CH$_2$-C**H**$_2$-CH$_2$-C≡C-H).

$^{13}$C{$^1$H} NMR (101 MHz): 191.44, 158.03, 135.02, 132.22, 132.06, 130.27, 129.48, 116.26, 112.75, 94.37, 80.59, 28.35, 25.05, 19.30, 14.20.

## 3.3 Preparation of 4-(6-(4’-bromophenoxy)hexyl)benzaldehyde (3)

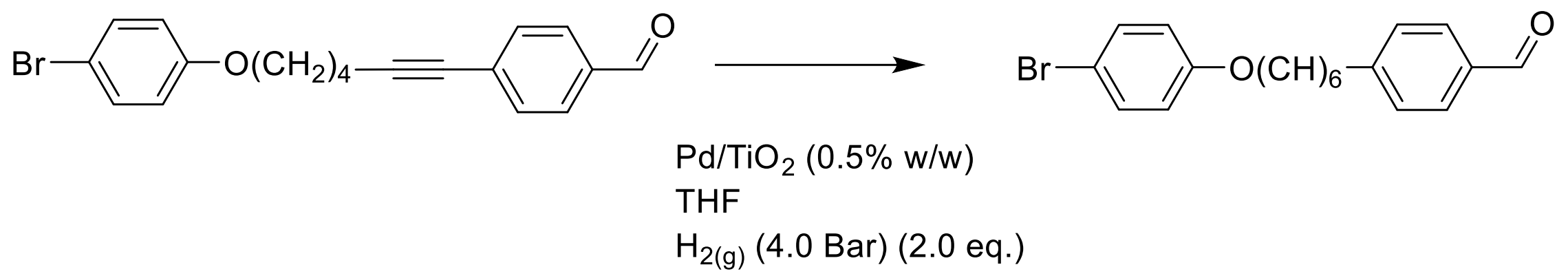

**Scheme S3.** Synthesis of 4-(6-(4’-bromophenoxy)hexyl)benzaldehyde **(4)**

The synthesis of the 4-(6-(4’-bromophenoxy)hexyl)benzaldehyde (**3**) is shown in **Scheme S3**. (**2**) (4.83 g, 15 mmol, 1eq.) was dissolved in 100 mL of dry THF. The dissolved sample was then paced into a high pressure bomb and fitted to a PARR 5500 Series Compact Reactor (HPCL) which was first sparged with $N_{2(g)}$ for 10 mins before $Pd/TiO_2$ (0.5%) (9.6 mg, 0.02% w/w) was added. The vessel was then put under 4 bar of $H_{2(g)}$, heated to 40 °C with constant stirring at 1200 rpm until 1.6 bar of $H_{2(g)}$ had been consumed (circa. 4 h). The crude product was then filtered through a pad of celite using $CHCl_3$ as the eluent. The crude product of 4-(6-(4’-bromophenoxy)hexyl)benzaldehyde was then retrieved as a colourless oil under reduced pressure and taken forward to the next reaction with no further purification.

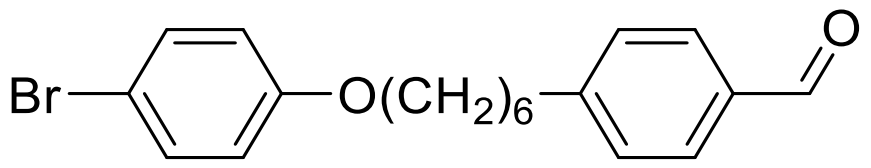

**(3)**

4-(6-(4’-bromophenoxy)hexyl)benzaldehyde

Yield: (Colourless oil) 5.12 g, 94 %.

## 3.4 Preparation of 4'-((6-(4-formylphenyl)hexyl)oxy)-[1,1'-biphenyl]-4-carbonitrile (4)

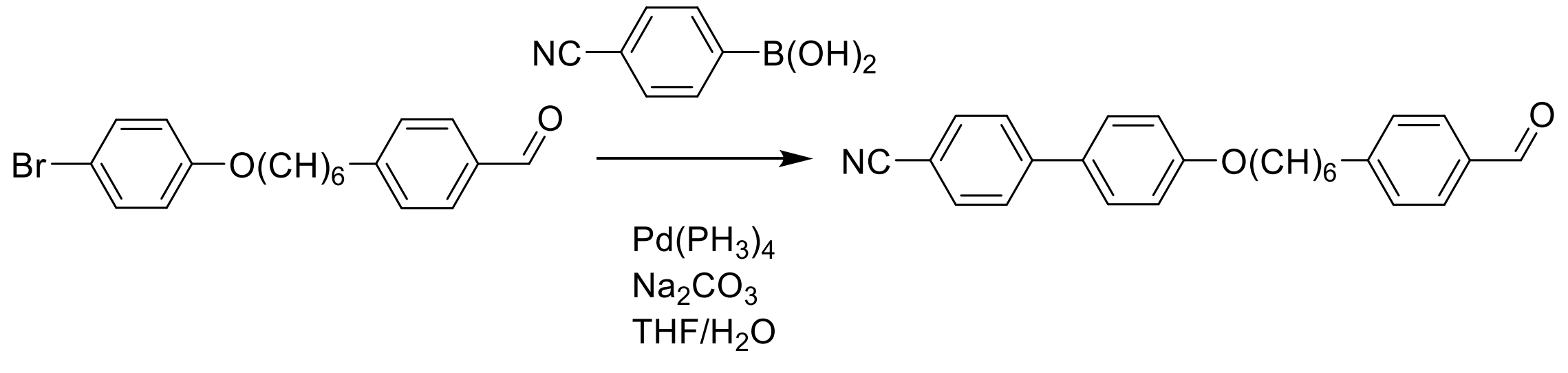

**Scheme S4.** Synthesis of 4'-((6-(4-formylphenyl)hexyl)oxy)-[1,1'-biphenyl]-4-carbonitrile

The synthesis of the 4'-((6-(4-formylphenyl)hexyl)oxy)-[1,1'-biphenyl]-4-carbonitrile (**4**) is shown in **Scheme S4**. (**3**) (3.0 g, 8.3 mmol, 1 eq.) and 4-cyanoboronic acid (1.37 g, 9.1 mmol, 1.1 eq.) was dissolved in 50 mL of THF. $Na_2CO_3$ (3.0 g, 30 mmol, 3.5 eq.) dissolved in 10 mL of water was then added and the reaction sparged with $Ar_{(g)}$ for 30 min. Tetrakis(triphenylphosphine)palladium(0) (7.5 mg, 0.025 eq.) was then added and the reaction heated to 60 °C for 12 h. The reaction was then cooled to room temperature, with the solvent being removed under reduced pressure. The crude sample was then dissolved in 25 mL of DCM and washed 2x 10 mL $Na_2CO_3$ and 1x 10 mL of brine. The crude sample was then purified further by column chromatography using DCM as the eluent. The off-white solid of 4'-((6-(4-formylphenyl)hexyl)oxy)-[1,1'-biphenyl]-4-carbonitrile was the retrieved under reduced pressure and re-crystallised using EtOH. The white powder recovered was then dried under reduced pressure overnight to give the reported yield.

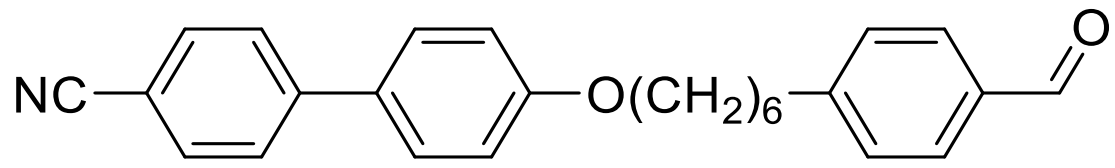

**(4)**

*4'-((6-(4-formylphenyl)hexyl)oxy)-[1,1'-biphenyl]-4-carbonitrile*

| | |
|---|---|
| Yield: | (white solid) 2.45 g, 77 %. |
| M.P.: | 91.0 °C |
| $R_f$ (DCM): | 0.33 |
| $^1$H NMR (400 MHz): | 9.97 (s, 1H, Ar-CO**H**), 7.80 (dd, *J* = 8.0, 1.1 Hz, 2H, Ar-**H**), 7.72 – 7.60 (m, 4H, Ar-**H**), 7.56 – 7.49 (m, 2H, Ar-**H**), 7.34 (dd, *J* = 7.9, 1.8 Hz, 2H, Ar-**H**), 6.98 (dd, *J* = 8.8, 1.7 Hz, 2H, Ar-**H**), 4.00 (t, *J* = 6.4 Hz, 2H, Ar-O-C$\mathbf{H_2}$), 2.71 (t, *J* = 7.7 Hz, 2H, $CH_2$-C$\mathbf{H_2}$-Ar), 1.82 (p, *J* = 7.8 Hz, 2H, Ar-O-$CH_2$-C$\mathbf{H_2}$-$CH_2$), 1.70 (p, *J* = 7.4 Hz, 2H, $CH_2$-C$\mathbf{H_2}$-$CH_2$-Ar), 1.56 – 1.47 (m, 2H, O-$CH_2$-$CH_2$-C$\mathbf{H_2}$-$CH_2$), 1.42 (p, *J* = 7.6 Hz, 2H, $CH_2$-C$\mathbf{H_2}$-$CH_2$-$CH_2$-Ar). |
| $^{13}$C{$^1$H} NMR (101 MHz): | 192.00, 159.72, 150.16, 145.25, 134.47, 132.58, 131.35, 130.16, 129.93, 129.70, 129.52, 129.26, 128.36, 126.40, 119.11, 67.98, 36.10, 30.95, 29.11, 28.92, 25.58. |

## 3.5 Preparation of the CBO6.O*m* series

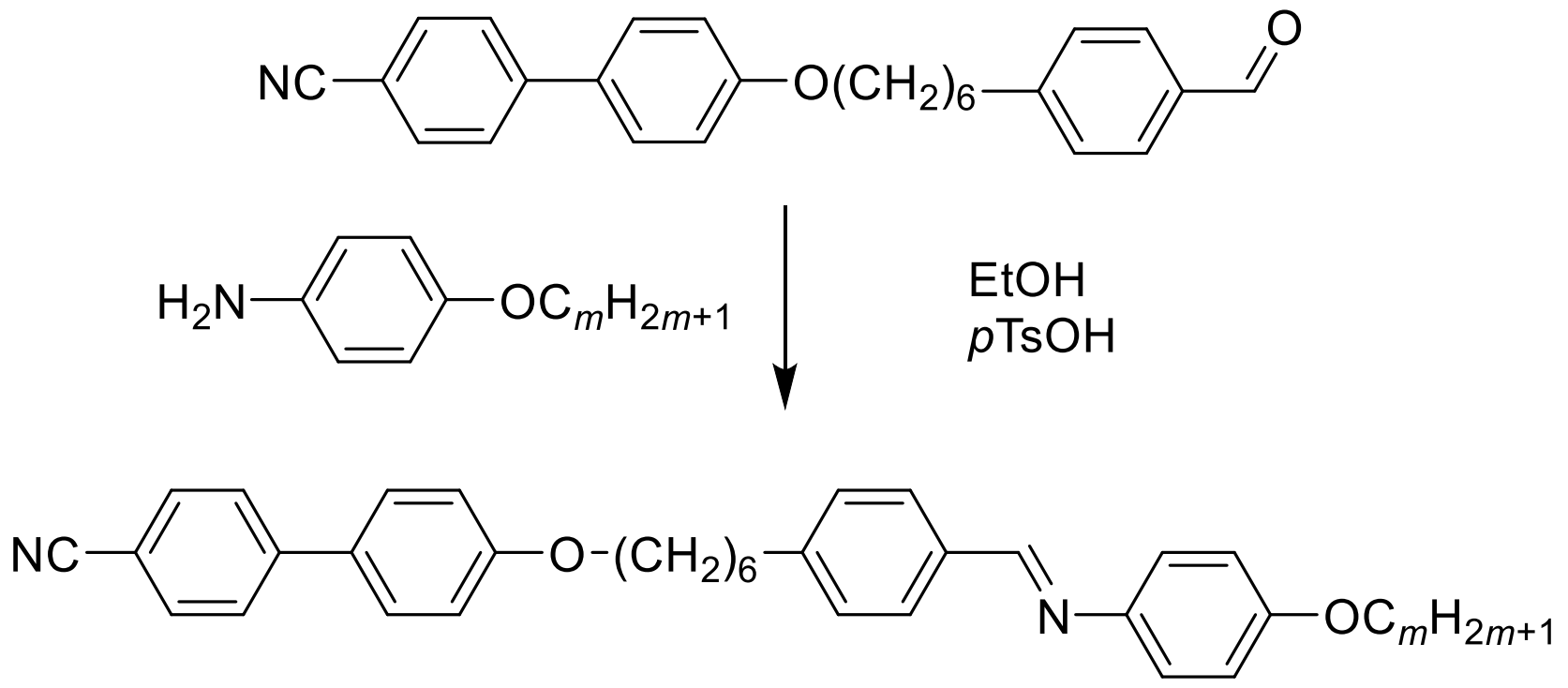

**Scheme S5**. Synthesis of 1-(4-cyanobiphenyl-4′-yloxy)-6-(4-alkyloxyanilinebenzylidene)-4′hexanes (CBO6.O*m*)

The synthesis of the CBO6.O*m* series is shown in **Scheme S5**. (**4**) (1 eq.) was dissolved in 10 mL ethanol at 60 °C. The relevant 4-alkoxyaniline (1.1 eq.) was then added followed by a single catalytic crystal of *p*TsOH. The quantities of the reactants used are show in **Table S2**. The reaction was then heated to 90 °C with constant stirring for 4 h before the crude products were retrieved by vacuum filtration. The crude products were then re-crystallised from EtOH. The final products of the 1-(4-cyanobiphenyl-4′-yloxy)-6-(4-alkyloxyanilinebenzylidene)-4′hexanes (CBO6.0*m*) were dried in a vacuum oven at 40 °C overnight to give the reported yields.

**Table S2.** Quantities of reagents used in the synthesis of the 1-(4-cyanobiphenyl-4′-yloxy)-6-(4-alkyloxyanilinebenzylidene)-4′hexanes (CBO6.0m)

| *CBO6.0m* | *Alkoxyaniline* | Quantity of *alkoxy aniline* | | | Quantity of I6.4.6 | |
|---|---|---|---|---|---|---|
| | | mg | mmol | µL | mg | mmol |
| *CBO6.02* | 4-Ethylaniline | 39.0 | 2.86 | 37.0 | 100.0 | 2.61 |
| *CBO6.03* | 4-Propylaniline | 43.0 | 2.86 | 42.0 | 100.0 | 2.61 |
| *CBO6.04* | 4-Butylaniline | 47.0 | 2.86 | 48.0 | 100.0 | 2.61 |
| *CBO6.06* | 4-Hexylaniline | 55.0 | 2.86 | - | 100.0 | 2.61 |
| *CBO6.07* | 4-Hepylaniline | 54.0 | 2.61 | - | 91.0 | 2.37 |
| *CBO6.08* | 4-Octylaniline | 63.0 | 2.86 | - | 100.0 | 2.61 |
| *CBO6.09* | 4-Nonylaniline | 52.0 | 2.21 | - | 77.0 | 2.01 |
| *CBO6.010* | 4-Decylaniline | 72.0 | 2.86 | - | 100.0 | 2.61 |

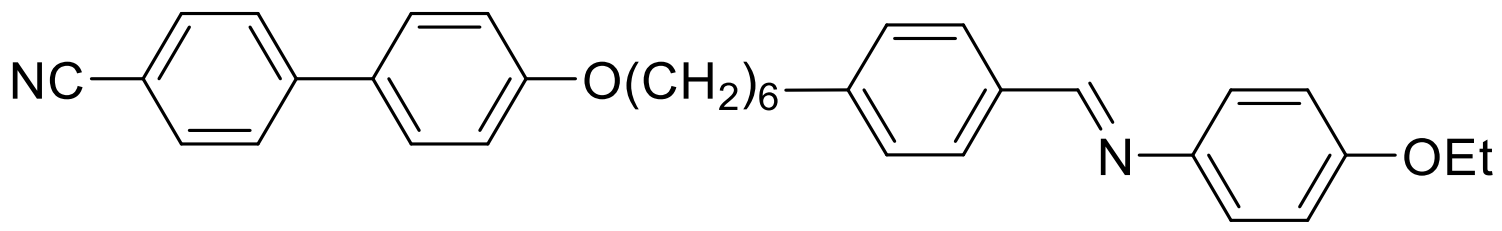

CBO6.O2

*1-(4-cyanobiphenyl-4'-yloxy)-6-(4-ethyloxyanilinebenzylidene)-4'hexane*

| | |
|---|---|
| Yield: | (white solid) 97 mg, 74 %. |
| $^1$H NMR (400 MHz): | 8.44 (s, 1H, Ar-C**H**=N), 7.79 (dd, *J* = 8.0, 1.6 Hz, 2H, Ar-**H**), 7.73 – 7.60 (m, 4H, Ar-**H**), 7.52 (dd, *J* = 8.7, 2.5 Hz, 2H, Ar-**H**), 7.28 (d, *J* = 2.1 Hz, 2H, Ar-**H**), 7.20 (dd, *J* = 8.7, 1.9 Hz, 2H, Ar-**H**), 6.98 (dd, *J* = 8.8, 2.0 Hz, 2H, Ar-**H**), 6.92 (dd, *J* = 8.8, 2.3 Hz, 2H, Ar-**H**), 4.06 (t, *J* = 7.0 Hz, 2H, Ar-O-C**H$_2$**), 3.99 (t, *J* = 6.4 Hz, 2H, Ar-O-C**H$_2$**-CH$_3$), 2.69 (t, *J* = 7.6 Hz, 2H, CH$_2$-C**H$_2$**-Ar), 1.81 (p, *J* = 6.7 Hz, 2H, O-CH$_2$-C**H$_2$**-(CH$_2$)$_2$, 1.70 (p, *J* = 7.5 Hz, 2H, (CH$_2$)$_2$-C**H$_2$**-CH$_2$-Ar), 1.55 – 1.46 (m, 3H, O-CH$_2$-C**H$_3$**), 1.46 – 1.38 (m, 4H, O-CH$_2$-CH$_2$-(C**H$_2$**)$_2$). |
| $^{13}$C{$^1$H} NMR (101 MHz): | 159.74, 158.31, 157.53, 146.14, 144.96, 134.21, 132.56, 128.85, 128.61, 128.33, 127.08, 122.12, 119.12, 115.09, 114.95, 68.03, 63.71, 35.85, 31.10, 29.09, 28.84, 25.88, 14.89. |
| M/Z [M+H]$^{+}$: | Calculated Values: $C_{34}H_{37}N_2O_2$; 503.2699. Found Values: 503.2681; Difference: 3.6 ppm. |

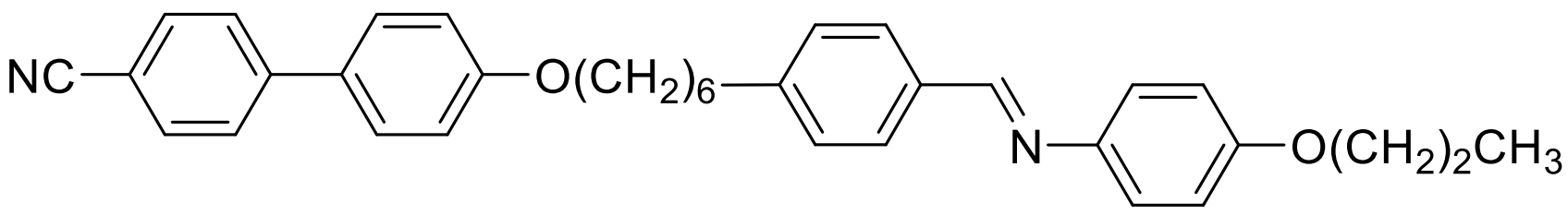

CBO6.03

*1-(4-cyanobiphenyl-4′-yloxy)-6-(4-propyloxyanilinebenzylidene)-4′hexane*

Yield: (white solid) 96 mg, 71 %.

$^{1}$H NMR (400 MHz): 8.45 (s, 1H, Ar-C**H**=N), 7.79 (dd, *J* = 8.0, 1.8 Hz, 2H, Ar-**H**), 7.72 – 7.60 (m, 4H, Ar-**H**), 7.52 (dd, *J* = 8.9, 2.8 Hz, 2H, Ar-**H**), 7.30 – 7.26 (m, 2H, Ar-**H**), 7.21 (dd, *J* = 8.8, 1.7 Hz, 2H, Ar-**H**), 7.03 – 6.94 (m, 2H, Ar-**H**), 6.92 (dd, *J* = 8.9, 2.3 Hz, 2H, Ar-**H**), 4.03 – 3.92 (m, 4H, Ar-O-C**H$_2$** and Ar-O-C**H$_2$**)*, 2.69 (t, *J* = 7.6 Hz, 2H, CH$_2$-C**H$_2$**-Ar), 1.88 – 1.75 (m, 4H, O-CH$_2$-C**H$_2$** and O-CH$_2$-C**H$_2$**)*, 1.71 (p, *J* = 7.5 Hz, 2H, (CH$_2$)$_2$-C**H$_2$**-CH$_2$-Ar), 1.54 – 1.48 (m, 4H, CH$_2$-(C**H$_2$**)$_2$-CH$_2$), 1.05 (t, *J* = 7.4 Hz, 3H, CH$_2$-C**H$_3$**).

$^{13}$C{$^{1}$H} NMR (101 MHz): 159.75, 158.26, 157.75, 146.12, 145.27, 134.22, 132.56, 129.09, 128.85, 128.61, 128.33, 127.08, 126.25, 122.17, 122.10, 119.12, 115.09, 114.98, 110.04, 69.80, 68.03, 35.85, 31.10, 29.09, 28.84, 25.88, 22.64, 10.55.

M/Z [M+H]$^{+}$: Calculated Values: $C_{35}H_{37}N_2O_2$; 517.2855. Found Values: 517.2834; Difference: 4.1 ppm.

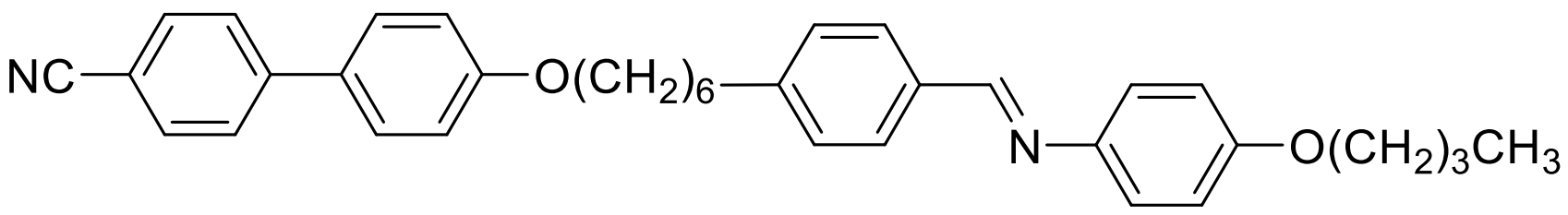

CBO6.04

*1-(4-cyanobiphenyl-4'-yloxy)-6-(4-butyloxyanilinebenzylidene)-4'hexane*

| | |
|---|---|
| Yield: | (white solid) 105 mg, 76 %. |
| $^{1}$H NMR (400 MHz): | 8.45 (s, 1H, Ar-C**H**=N), 7.80 (dd, *J* = 7.9, 1.6 Hz, 2H, Ar-**H**), 7.72 – 7.59 (m, 4H, Ar-**H**), 7.52 (dd, *J* = 8.8, 2.0 Hz, 2H, Ar-**H**), 7.28 (m, 2H, Ar-**H**), 7.20 (dd, *J* = 8.9, 1.9 Hz, 2H, Ar-**H**), 6.98 (dd, *J* = 8.8, 1.7 Hz, 2H, Ar-**H**), 6.92 (dd, *J* = 8.8, 1.7 Hz, 2H, Ar-**H**), 4.11 – 3.96 (m, 4H, Ar-O-C$\mathbf{H_2}$ and Ar-O-C$\mathbf{H_2}$)*, 2.69 (t, *J* = 7.6 Hz, 2H, $CH_2$-C$\mathbf{H_2}$-Ar), 1.81 (p, *J* = 6.7 Hz, 2H, O-$CH_2$-C$\mathbf{H_2}$), 1.70 (p, *J* = 7.6 Hz, 2H, O-$CH_2$-C$\mathbf{H_2}$), 1.56 (m, 6H, $CH_2$-(C$\mathbf{H_2}$)$_2$-$CH_2$ and O-($CH_2$)$_2$-C$\mathbf{H_2}$-$CH_3$)*, 1.43 (t, *J* = 7.0 Hz, 3H, $CH_2$-C$\mathbf{H_3}$). |
| $^{13}$C{$^{1}$H} NMR (101 MHz): | 159.75, 158.30, 157.53, 146.13, 145.27, 144.97, 134.22, 132.56, 131.30, 128.85, 128.61, 128.33, 127.08, 122.19, 122.12, 119.12, 115.09, 114.95, 110.04, 68.03, 63.71, 35.85, 31.10, 29.09, 28.84, 25.88, 14.90. |
| M/Z $[M+H]^{+}$: | Calculated Values: $C_{36}H_{39}N_2O_2$; 531.3012. Found Values: 531.3004; Difference: 1.5 ppm. |

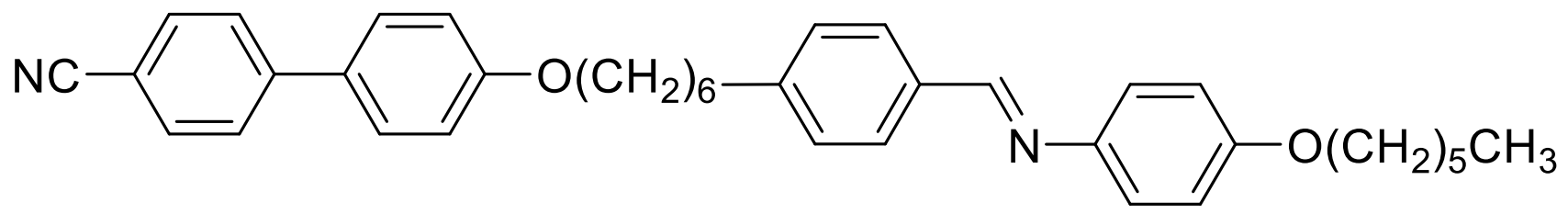

CBO6.06

*1-(4-cyanobiphenyl-4'-yloxy)-6-(4-hexyloxyanilinebenzylidene)-4'hexane*

Yield: (white solid) 105 mg, 76 %.

$^{1}$H NMR (400 MHz): 8.45 (s, 1H, Ar-C**H**=N), 7.79 (dd, *J* = 7.9, 2.0 Hz, 2H, Ar-**H**), 7.72 – 7.59 (m, 4H, Ar-**H**), 7.52 (dd, *J* = 8.2, 1.5 Hz, 2H, Ar-**H**), 7.21 (dd, *J* = 8.3, 1.5 Hz, 2H, Ar-**H**), 7.06 – 6.86 (m, 4H, Ar-**H**), 4.06 – 3.91 (m, 4H, Ar-O-C$\mathbf{H_2}$ and Ar-O-C$\mathbf{H_2}$)*, 2.69 (t, *J* = 7.6 Hz, 2H, $CH_2$-C$\mathbf{H_2}$-Ar), 1.89 – 1.75 (m, 4H, O-$CH_2$-C$\mathbf{H_2}$ and O-$CH_2$-C$\mathbf{H_2}$)*, 1.70 (p, *J* = 7.6 Hz, 2H, $(CH_2)_2$-C$\mathbf{H_2}$-$CH_2$-Ar)1.54 – 1.26 (m, 10H, $CH_2$-(C$\mathbf{H_2}$)$_2$-$CH_2$ and O-$(CH_2)_2$-(C$\mathbf{H_2}$)$_3$-$CH_3$)*, 0.91 (t, *J* = 6.7 Hz, 3H, $CH_2$-C$\mathbf{H_3}$).

$^{13}$C{$^{1}$H} NMR (101 MHz): 159.75, 158.24, 134.24, 132.56, 131.30, 128.84, 128.60, 128.33, 127.08, 122.10, 119.12, 118.42, 115.54, 115.09, 114.97, 114.97, 68.31, 68.03, 35.86, 31.62, 31.11, 29.30, 28.85, 25.89, 25.75, 22.63, 14.06.

M/Z $[M+H]^{+}$: Calculated Values: $C_{38}H_{43}N_2O_2$; 559.3325. Found Values: 559.3325; Difference: 0.0 ppm.

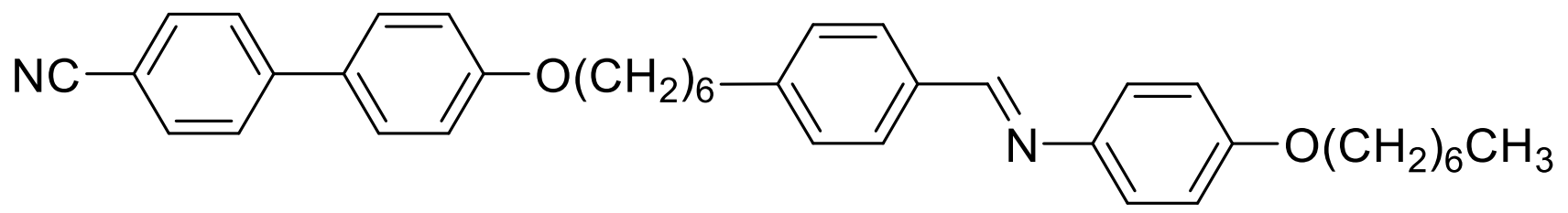

CBO6.07

*1-(4-cyanobiphenyl-4′-yloxy)-6-(4-heptyloxyanilinebenzylidene)-4′hexane*

Yield: (white solid) 96 mg, 71 %.

$^1$H NMR (400 MHz): 8.45 (s, 1H, Ar-C**H**=N), 7.79 (dd, *J* = 8.1, 1.7 Hz, 2H, Ar-**H**), 7.72 – 7.60 (m, 4H, Ar-**H**), 7.52 (dd, *J* = 8.7, 2.1 Hz, 2H, Ar-**H**), 7.31 – 7.26 (m, 2H, Ar-**H**), 7.20 (dd, *J* = 8.7, 2.0 Hz, 2H, Ar-**H**), 6.98 (dd, *J* = 8.8, 1.9 Hz, 2H, Ar-**H**), 6.91 (dd, *J* = 8.9, 2.0 Hz, 2H, Ar-**H**), 4.03 – 3.94 (m, 4H, Ar-O-C**H$_2$** and Ar-O-C**H$_2$**)*, 2.69 (t, *J* = 7.6 Hz, 2H, CH$_2$-C**H$_2$**-Ar), 1.80 (m, 4H, O-CH$_2$-C**H$_2$** and O-CH$_2$-C**H$_2$**)*, 1.70 (p, *J* = 7.6 Hz, 2H, (CH$_2$)$_2$-C**H$_2$**-CH$_2$-Ar), 1.29 (m, 12H, CH$_2$-(C**H$_2$**)$_2$-CH$_2$ and O-(CH$_2$)$_2$-(C**H$_2$**)$_4$-CH$_3$)*, 0.88 (t, *J* = 7.0, 6.5 Hz, 3H, CH$_2$-C**H$_3$**).

$^{13}$C{$^1$H} NMR (101 MHz): 158.23, 157.76, 146.11, 144.90, 134.24, 132.56, 131.30, 128.84, 128.60, 128.33, 127.08, 122.10, 119.12, 115.09, 114.97, 110.05, 68.32, 68.03, 35.85, 31.89, 31.11, 29.56, 29.43, 29.34, 29.28, 29.10, 28.85, 26.07, 25.89, 22.69, 14.13.

M/Z [M+H]$^+$: Calculated Values: $C_{39}H_{45}N_2O_2$; 573.3481. Found Values: 573.3456; Difference: 4.4 ppm.

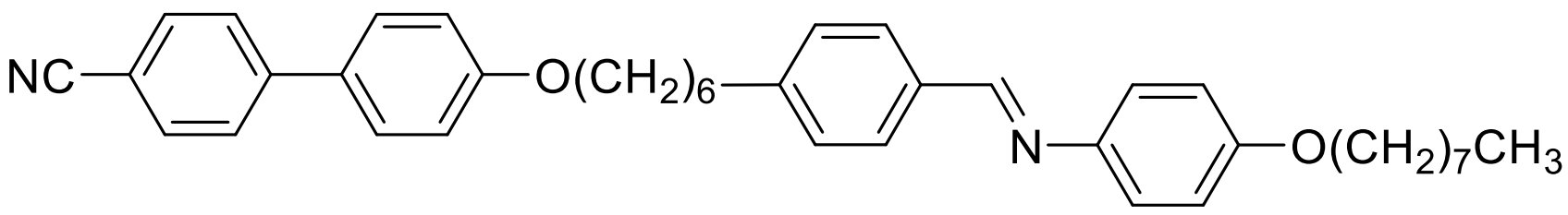

CBO6.08

*1-(4-cyanobiphenyl-4'-yloxy)-6-(4-octyloxyanilinebenzylidene)-4'hexane*

Yield: (white solid) 61 mg, 40 %.

$^1$H NMR (400 MHz): 8.45 (s, 1H, Ar-C**H**=N), 7.79 (dd, *J* = 8.0, 1.1 Hz, 2H, Ar-**H**), 7.72 – 7.60 (m, 4H, Ar-**H**), 7.52 (dd, *J* = 9.0, 2.2 Hz, 2H, Ar-**H**), 7.30 – 7.26 (m, 2H, Ar-**H**), 7.20 (t, *J* = 8.9, 1.7 Hz, 2H, Ar-**H**), 6.99 (dd, *J* = 8.7, 1.7 Hz, 2H, Ar-**H**), 6.92 (dd, *J* = 8.9, 2.1 Hz, 2H, Ar-**H**), 4.02 – 3.93 (m, 4H, Ar-O-C**H₂** and Ar-O-C**H₂**)*, 2.69 (t, *J* = 7.6 Hz, 2H, $CH_2$-C**H₂**-Ar), 1.86 – 1.74 (m, 4H, O-$CH_2$-C**H₂** and O-$CH_2$-C**H₂**)*, 1.70 (p, *J* = 7.6 Hz, 2H,($CH_2$)$_2$-C**H₂**-$CH_2$-Ar), 1.54 – 1.21 (m, 14H, $CH_2$-(C**H₂**)$_2$-$CH_2$ and O-($CH_2$)$_2$-(C**H₂**)$_5$-$CH_3$), 0.89 (t, *J* = 6.6 Hz, 3H, $CH_2$-C**H₃**).

$^{13}$C{$^1$H} NMR (101 MHz): 159.75, 158.24, 145.27, 144.90, 132.56, 132.56, 131.30, 128.84, 128.60, 128.33, 127.08, 122.10, 119.12, 115.09, 114.97, 68.32, 68.03, 35.85, 31.83, 31.11, 29.39, 29.33, 29.26, 29.10, 28.85, 26.08, 25.89, 22.68, 14.12.

M/Z [M+H]$^+$: Calculated Values: $C_{41}H_{47}N_2O_2$; 589.3423. Found Values: 589.1221; Difference: 4.1 ppm.

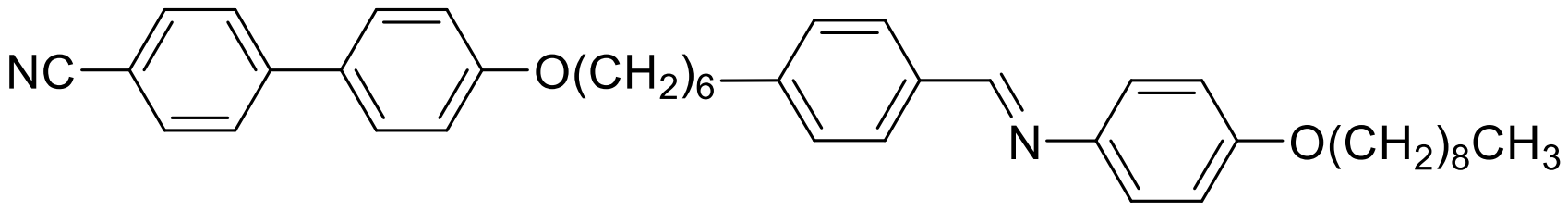

CBO6.09

*1-(4-cyanobiphenyl-4'-yloxy)-6-(4-nonyloxyanilinebenzylidene)-4'hexane*

Yield: (white solid) 98 mg, 81 %.

$^{1}$H NMR (400 MHz): 8.45 (s, 1H, Ar-C**H**=N), 7.79 (dd, *J* = 8.0, 1.6 Hz, 2H, Ar-**H**), 7.72 – 7.60 (m, 4H, Ar-**H**), 7.52 (dd, *J* = 8.9, 2.7 Hz, 2H, Ar-**H**), 7.30 – 7.26 (m, 2H, Ar-**H**), 7.20 (dd, *J* = 8.9, 2.0 Hz, 2H, Ar-**H**), 6.99 (dd, *J* = 8.9, 2.0 Hz, 2H, Ar-**H**), 6.92 (dd, *J* = 8.9, 2.0 Hz, 2H, Ar-**H**), 4.03 – 3.91 (m, 4H, Ar-O-C**H₂** and Ar-O-C**H₂**)*, 2.69 (t, *J* = 7.6 Hz, 2H, $CH_2$-C**H₂**-Ar), 1.88 – 1.74 (m, 4H, O-$CH_2$-C**H₂** and O-$CH_2$-C**H₂**)*, 1.70 (p, *J* = 7.6 Hz, 2H,$(CH_2)_2$-C**H₂**-$CH_2$-Ar), 1.46 – 1.24 (m, 16H, $CH_2$-(C**H₂**)$_2$-$CH_2$ and O-$(CH_2)_2$-(C**H₂**)$_6$-$CH_3$)*, 0.90 (t, *J* = 7.2, 6.5 Hz, 3H, $CH_2$-C**H₃**).

$^{13}$C{$^{1}$H} NMR (101 MHz): 159.64, 158.77, 158.24, 151.05, 150.52, 149.93, 146.11, 145.27, 144.90, 143.72, 141.86, 141.43, 137.93, 132.56, 128.84, 128.60, 128.33, 127.08, 122.16, 122.10, 119.12, 115.09, 114.97, 68.31, 68.03, 35.85, 31.81, 31.11, 29.34, 29.10, 26.04, 22.63, 14.11.

M/Z $[M+H]^+$: Calculated Values: $C_{41}H_{49}N_2O_2$; 601.3794. Found Values: 601.3782; Difference: 2.0 ppm.

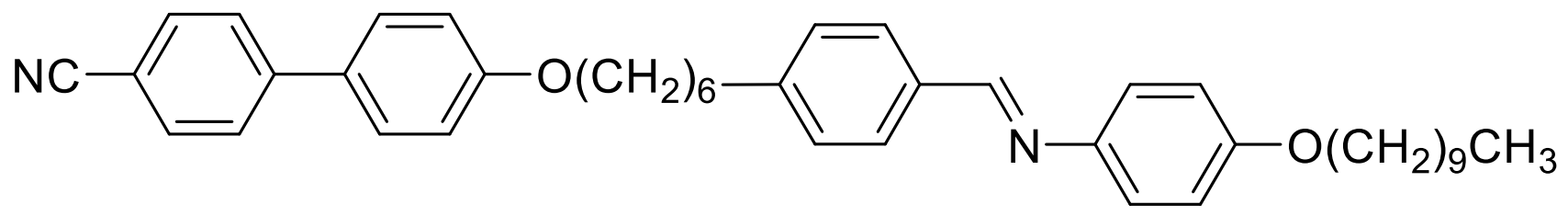

CBO6.010

*1-(4-cyanobiphenyl-4'-yloxy)-6-(4-decyloxyanilinebenzylidene)-4'hexane*

Yield: (white solid) 132 mg, 82 %.

$^1$H NMR (400 MHz): 8.44 (s, 1H, Ar-C**H**=N), 7.80 (dd, *J* = 7.9, 2.4 Hz, 2H, Ar-**H**), 7.71 – 7.59 (m, 4H, Ar-**H**), 7.52 (dd, *J* = 8.8, 2.2 Hz, 2H, Ar-**H**), 7.28 (m, 2H, Ar-**H**), 7.20 (dd, *J* = 8.8, 1.3 Hz, 2H, Ar-**H**), 6.99 (dd, *J* = 8.9, 2.6 Hz, 2H, Ar-**H**), 6.92 (dd, *J* = 8.8, 1.5 Hz, 2H, Ar-**H**), 4.02 – 3.89 (m, 4H, Ar-O-C**H$_2$** and Ar-O-C**H$_2$**)*, 2.69 (t, *J* = 7.6 Hz, 2H, $CH_2$-C**H$_2$**-Ar), 1.80 (m, 4H, O-$CH_2$-C**H$_2$** and O-$CH_2$-C**H$_2$**)*, 1.70 (p, *J* = 7.6 Hz, 2H, $(CH_2)_2$-C**H$_2$**-$CH_2$-Ar), 1.47 – 1.29 (m, 18H, $CH_2$-(C**H$_2$**)$_2$-$CH_2$ and O-$(CH_2)_2$-(C**H$_2$**)$_7$-$CH_3$)*, 0.89 (t, *J* = 6.6 Hz, 3H, $CH_2$-C**H$_3$**).

$^{13}$C{$^1$H} NMR (101 MHz): 159.75, 158.24, 157.76, 146.12, 145.27, 144.89, 134.23, 132.56, 131.30, 128.84, 128.60, 128.33, 127.08, 122.16, 122.10, 119.12, 115.09, 114.99, 114.97, 110.05, 68.32, 68.03, 35.85, 31.91, 31.11, 29.60, 29.58, 29.43, 29.34, 29.10, 28.86, 26.07, 25.89, 22.70, 14.13.

M/Z $[M+H]^+$: Calculated Values: $C_{42}H_{51}N_2O_2$; 615.3951. Found Values: 615.3933; Difference: 2.9 ppm